\documentclass[prl,aps,amssymb,twocolumn,superscriptaddress,notitlepage]{revtex4-2}
\usepackage{amsmath}
\usepackage{amssymb}
\usepackage{amsthm}
\usepackage{amsfonts}
\usepackage{listings}
\usepackage{physics}
\usepackage{enumerate}
\usepackage{latexsym}
\usepackage{psfrag}
\usepackage{bm}
\usepackage{graphicx}
\usepackage[caption=false]{subfig}
\usepackage{blkarray}
\usepackage{array}
\usepackage{color}
\usepackage[normalem]{ulem}
\usepackage{hyperref}
\hypersetup{
     colorlinks = true,
     linkcolor = magenta,
     citecolor = magenta
}
\usepackage{mathtools}

\renewcommand\rm{\mathrm}

\newcommand\e{\epsilon}

\renewcommand\dd{\mathrm{d}}
\newcommand{\Vi}{\mathsf{Vi}}
\newcommand{\Ch}{\mathsf{Ch}}
\newcommand{\xm}[1] {{#1}}
\newcommand{\nb}[1] {{#1}}
\renewcommand{\ss}[1] {{#1}}

\allowdisplaybreaks

\begin{document}
\title{Mechanochemical feedback drives complex inertial dynamics in active solids
} 
\author{Siddhartha Sarkar}
\thanks{These two authors contributed equally.}
\affiliation{Department of Physics, University of Michigan, Ann Arbor, MI 48109, USA}
\affiliation{Max Planck Institute for the Physics of Complex Systems, N\"othnitzer Stra\ss e 38, 01187 Dresden, Germany}

\author{Biswarup Ash}
\thanks{These two authors contributed equally.}
\affiliation{Department of Physics, University of Michigan, Ann Arbor, MI 48109, USA}

\author{Yueyang Wu}
\affiliation{Department of Physics, University of Michigan, Ann Arbor, MI 48109, USA}

\author{Nicholas Boechler}
\email{nboechler@ucsd.edu}
\affiliation{Department of Mechanical and Aerospace Engineering, University of California San Diego, La Jolla, CA 92093, USA}
\affiliation{Program in Materials Science and Engineering, University of California San Diego, La Jolla, CA 92093, USA}

\author{Suraj Shankar}
\email{surajsh@umich.edu}
\affiliation{Department of Physics, University of Michigan, Ann Arbor, MI 48109, USA}

\author{Xiaoming Mao}
\email{maox@umich.edu}
\affiliation{Department of Physics, University of Michigan, Ann Arbor, MI 48109, USA}

\begin{abstract} 
Active solids combine internal active driving 
with elasticity to realize states with nonequilibrium mechanics and autonomous motion. They are often studied in overdamped settings, e.g., in soft materials, and the role of inertia is less explored.  We construct a model of a chemically active solid that incorporates mechanochemical feedback and show that, when feedback overwhelms mechanical damping, autonomous inertial dynamics can spontaneously emerge through sustained consumption of chemical fuel. By combining numerical simulations, analysis and dynamical systems approaches, we show how active feedback drives complex nonlinear dynamics on multiple time-scales, including limit cycles and chaos. Our results suggest design principles for creating ultrafast actuators and autonomous machines from soft, chemically-powered solids. 
\end{abstract}
\maketitle




\noindent{\it Introduction.}
Soft materials driven by chemical stimuli have been widely used as actuators to achieve complex tasks---locomotion, grasping, and shape-morphing---through flexible deformations \cite{asaka2019soft,rus2015design}. While promising for technologies \cite{li2022soft}, their contraction speed and power output are often limited by slow, diffusive mechanisms (thermal, fluidic, ionic, etc.) and damping~\cite{de1997semi,mirvakili2018artificial}. To overcome this limitation, recent approaches have used miniaturization~\cite{he2021electrospun,bas2021ultrafast} or inertial dynamics triggered by plant-inspired elastic instabilities~\cite{skotheim2005physical,lee2010first,gorissen2020inflatable} and fast chemical processes (e.g.,  combustion)~\cite{chen2019controlled,mao2022ultrafast,shepherd2013using,aubin2023powerful} to achieve fast actuation. However, such systems passively relax without sustained input, prompting us to ask: How can soft, chemically driven materials sustain fast and continuous inertial motion without cyclic stimulation, and how complex can this motion be?

A potential strategy is suggested by contractile tissues that use biological feedback to drive extreme motions in animals \cite{armon2018ultrafast,shankar2024active}. Such active solids, 
exemplified by living tissues \cite{banerjee2019continuum,noll2017active,boocock2021theory,shankar2024active}, cellular and organismal collectives \cite{tan2022odd,xu2023autonomous},  mechanochemical gels \cite{yashin2006pattern,yashin2012mechano,levin2020self} and robotic assemblies \cite{baconnier2022selective}, have been shown to produce exotic dynamics~\cite{scheibner2020odd,banerjee2015propagating,boocock2021theory,maitra2019oriented,fruchart2023odd,noll2017active,tan2022odd,shankar2024active} and present new opportunities for designing soft chemically-driven machines.

Building upon previous studies on biological systems with overdamped feedback \cite{banerjee2015propagating,notbohm2016cellular,banerjee2017actomyosin,boocock2021theory,parmar2025spontaneous,shankar2024active,chao2024selective} and inertial robotic assemblies \cite{brandenbourger2019non,brandenbourger2021limit,veenstra2024non}, we combine both effects to demonstrate how chemically active soft solids with sufficiently strong mechanochemical feedback can sustain complex and autonomous nonlinear inertial dynamics. We show how feedback introduces new timescales that govern how stress accelerates or decelerates chemical reactions. When this response becomes faster than viscous dissipation of momentum, the system continually extracts chemical energy to power self-sustained inertial oscillations and chaotic dynamics. This suggests mechanochemical feedback as a promising scalable mechanism for devising soft engines, actuators, and autonomous machines with fast and rich dynamics.

\noindent{\it Mechanochemical active solid.}
We model the active solid as a one-dimensional (1D) gel of length $L$ and mass density $\rho$ suffused with a mixture of chemical species $A$ and $B$ (concentration $n_A, n_B$)  that inter-convert into each other via a reversible reaction $A\xrightleftharpoons[]{}B$. The total concentration is conserved by the reaction ($n_A+n_B=n_0$), so the relative concentration difference defined by $\chi=(n_B-n_A)/n_0$ serves as a convenient reaction coordinate. Material deformations are assumed small and captured by a linearized strain $\e(x,t)=\partial_xu(x,t)$ ($u$: displacement, $x$: material coordinate). The free energy of the system \cite{de2013non,chaikin1995principles,prost2015active,marchetti2013hydrodynamics},
\begin{equation}\label{EQ:FE}
    \mathcal{F}=k_BT\int\dd x\left[n_0f(\chi)+\dfrac{E}{2k_BT}\e^2+n_0C\chi\e\right]\;,
\end{equation}
combines a local free energy density $f(\chi)$, an elastic energy with modulus $E$ and a passive coupling $C$ between the chemical concentration and deformations, with $T$ the temperature. The passive coupling controls the chemically induced mechanical stresses present,  e.g., as can be seen in chemically responsive gels~\cite{yashin2006pattern,levin2020self}, solvent-driven swelling \cite{lee2010first}, and explosives or combustion \cite{aubin2023powerful,cooper1996explosives}. 

The mechanical stress in the material $\sigma=\sigma^p+\sigma^d$ consists of both passive ($\sigma^p=\delta\mathcal{F}/\delta\e=E\e+n_0k_BT C\chi$) and dissipative ($\sigma^d=\eta\partial_t\e$) contributions \cite{chaikin1995principles,prost2015active,marchetti2013hydrodynamics}, with $\eta$ a viscosity. In the absence of any reactive chemicals ($n_0=0$), the material behaves as a passive Kelvin-Voigt viscoelastic solid.
Kinematics and momentum conservation then gives
\begin{align}\label{eq:duchi}
    &\partial_tu=v\;,\nonumber\\
    &\rho\partial_tv=\partial_x(\sigma^p+\sigma^d) \;,
\end{align}
where $v(x,t)$ is the local material velocity.

Chemical reactions involve transitions between molecular states and, in equilibrium, a balance of forward and backward transition rates \cite{prigogine1958chemical,hill1986introduction}. That mechanical deformations can modify reaction rates (kinetics) and bias chemical equilibria (energetics) is known in diverse systems, e.g., mechanochemically responsive polymers \cite{hickenboth2007biasing,garcia2017steering,ghanem2021role}, inorganic compunds \cite{gilman1996mechanochemistry,o2021many}, biomolecules \cite{bell1978models,keller2000mechanochemistry,bustamante2004mechanical}, and living cells \cite{vogel2018unraveling,bailles2022mechanochemical}.

Following standard transition state theory~\cite{prigogine1958chemical,hanggi1990reaction} generalized for non-ideal systems~\cite{zwicker2022intertwined,aslyamov2023nonideal}, we have forward and backward reaction rates given by $r_+=k_0\exp(\mu_A/k_BT)$ and $r_-=k_0\exp(\mu_B/k_BT)$, respectively, where $\mu_A, \mu_B$ are the chemical potentials. Here $k_0$ is a microscopic rate constant that depends on details of the transition state. In ideal systems, these rates reduce to the law of mass action.
In a passive system, $\mu^p_i=\delta\mathcal{F}/\delta n_i$, 
so $\mu^{p}_B=-\mu_A^{p}=k_BT[f'(\chi)+C\e]$. This relation demonstrates reciprocity between the reaction and strain ($\delta\mu_{i}^{p}/\delta\e=\delta\sigma^p/\delta n_i=\mp k_B T C$, $i=A,B$) \cite{de2013non,hill1986introduction} and captures the energetic shift in chemical equilibria due to deformations.

To drive the system out of chemical equilibrium, we augment the reaction rates by setting $\mu_i=\mu_i^{p}+\mu^a_i$, where 
an external source of reactive free energy, say from an abundant supply of a fuel that catalyses the reaction \cite{prost2015active,marchetti2013hydrodynamics}, drives a bias in chemical potential. A constant bias simply changes the equilibrium concentration of reactants and products by a fixed amount and leaves the system passive. But a dynamic bias
can induce an active coupling between mechanical stress and the reaction beyond the free energy  in Eq.~\eqref{EQ:FE}.  In particular, 
mechanical stress can modify both the transition state barrier and the chemical potential difference, which we minimally capture by setting $\mu^a_i=h_i\sigma$, where $\sigma$ is the total stress and $h_{i}$ are mechanochemical coupling constants. This feedback breaks reciprocity as $\delta\mu_i/\delta\e\neq \delta\sigma^p/\delta n_i$,
and provides a nonequilibrium coupling between the reaction and mechanical degrees of freedom.

Collecting these effects together, and assuming  $\mu_i\ll k_BT$ (the chemical potential associated with the reaction is small) to linearize the reaction rates,  
we obtain (see more details in the Supplementary Materials (SM)~\cite{SM2024})
\begin{align}\label{eq:dchi}
\partial_t\chi=r_+-r_- = r_{\rm{eq}} + \zeta\sigma,
\end{align}
where $ r_{\rm eq} = (k_0/k_BT){(\mu^p_A-\mu^p_B)}=-2k_0[f'(\chi)+C\e]$ is the equilibrium reaction rate including the passive coupling to strain and $\zeta=k_0(h_A-h_B)/k_BT$ is the non-equilibrium coupling to stress. 
{
Note that for the system to be passive, we need $\zeta=0$, which invalidates coupling to strain rate.
}
Eqs.~(\ref{eq:duchi}-\ref{eq:dchi}) along with initial and boundary conditions completely specify the inertial active solid model. 

\begin{figure}
    \centering
    \includegraphics[width=0.49 \textwidth]{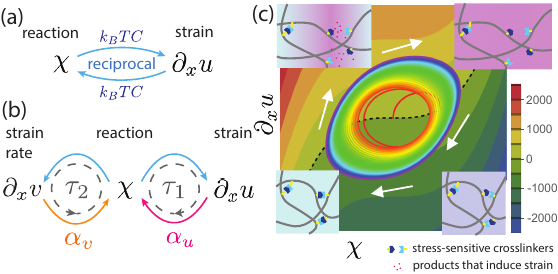}
    \caption{Mechanochemical coupling. 
     (a,b) Feedback loops among reaction ($\chi$), strain ($\partial_x u$), strain rate ($\partial_x v$) in {passive} (a) and active (b) solids. { In the passive limit (a), detailed balance enforces vanishing strain-rate coupling, and a reciprocal strain-reaction coupling ($k_BTC$). When active (b), the feedback drives nonequilibrium fluxes on independent time scales $\tau_{1,2}$.} 
     (c) Contour plot of reaction rate $\partial_t \chi$ (in A.U.) using the 0D model in Eq.~\eqref{eq:0D}, with the dashed line marking $\partial_t \chi=0$ so $\chi$ grows (decays) above (below) this curve. An example trajectory 
     is shown where a perturbation from the origin leads to a limit cycle, with four representative states illustrated as insets, \nb{where the background shading refers to concentration $\chi$. Starting with the bottom left corner, increased strain breaks stress-sensitive crosslinkers, which drives $\chi$ to increase, followed by decrease in strain, and recovery back to the first state.} 
     }
    \label{fig:barrier}
\end{figure}

\begin{figure*}
    \centering
    \includegraphics[width=\textwidth]{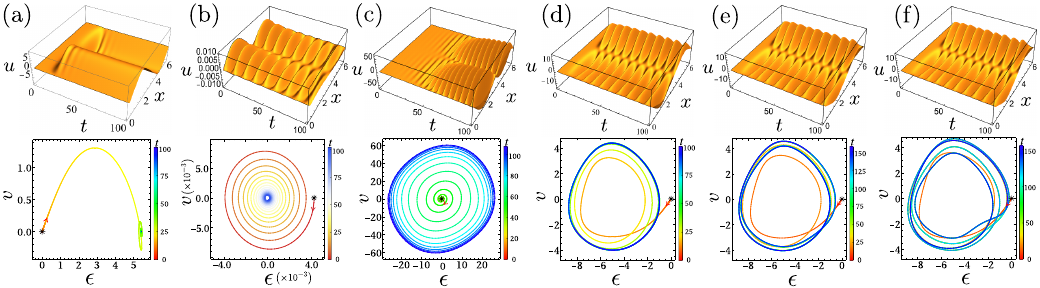}
    \caption{Dynamics of the 1D mechanochemical system in Eq.~\eqref{eq:1Deqs}. We set $L  = 2\pi$, $\eta = 0.16$, and $\tau_c = 100$ along with    
    (a) $\alpha_u = 50, \alpha_v = 0$, (b) $\alpha_u = -10, \alpha_v = 0$, (c) $\alpha_u = -50, \alpha_v = 0$, 
    (d) $\alpha_u = 50, \alpha_v = 77$, (e) $\alpha_u = 50, \alpha_v = 80$, (f) $\alpha_u = 50, \alpha_v = 82$.
    We include a small diffusion constant $D = 0.05$ to regularize chemical gradients. The initial conditions are $u(x,0) = 0.01\sin(x)$, $v(x,0) = 0 = \chi(x,0)$ 
    The first row shows the displacement field $u(x,t)$. The second row shows the phase space $({\epsilon=\partial_x u},v)$ trajectory at the point $x = 9\pi/25$. 
    }
    \label{fig:1D}
\end{figure*}

\noindent{\it 1D model.}
To exemplify its key consequences, we consider a bistable chemical free energy $f(\chi)=-a_2\chi^2/2+a_4\chi^4/4$ with $a_2,a_4>0$. The model (Eqs.~(\ref{eq:duchi}-\ref{eq:dchi})) then simplifies to give
\begin{align}
\label{eq:1Deqs}
    &\partial_tu=v\;,\nonumber\\
    &\rho\partial_t v=E\partial_x^2 u +\eta \partial_x^2 v + c\partial_x\chi,\\
  &\tau_c\partial_t\chi =\chi(1-g\chi^2) - \alpha_u \partial_x u-\alpha_v\partial_x v,\nonumber
\end{align}
\ss{where $c=n_0k_BTC$ is the passive chemically induced stress scale, $\tau_c=1/[2k_0a_2+\zeta c]$ is the effective chemical reaction timescale, combining the equilibrium rate ($2k_0a_2$) and the active feedback rate ($\zeta c$)~\cite{SM2024},} $g=2k_0a_4\tau_c$ controls the chemical nonlinearity, and $\alpha_u=\tau_c(2k_0C-\zeta E)$ and $\alpha_v=-\tau_c\zeta\eta$ are the nonequilibrium feedback couplings to strain and strain-rate, respectively \xm{(Fig.~\ref{fig:barrier}a,b)}. 

There are three homogeneous steady-state solutions, the trivial one ($\chi=\chi_{ss}, \e=\e_{ss}$, and $v=0$), which always exists, and two non-trivial ones ($\e_{ss}=\pm (c/E)\sqrt{(E+c\alpha_u)/Eg}$, $\chi_{ss}=\mp\sqrt{(E+c\alpha_u)/Eg}$, and $v=0$), which only exist for $c\alpha_u<-E$.  
We numerically solve Eqs.~\eqref{eq:1Deqs} in a 1D domain of length $L=2\pi$ with periodic boundary conditions using the pseudo-spectral method~\cite{orszag1972comparison}. 
We pick time, length, mass, and reaction coordinate scales so that $\rho=E=c=g=1$ and vary the activity parameters, $\alpha_u$ and $\alpha_v$. 

In Fig.~\ref{fig:1D}, we showcase rich dynamics that can emerge from this 1D model. 
For $\alpha_v=0$ (only strain feeding back to reaction), as $\alpha_u$ decreases from positive (reciprocal) to negative (nonreciprocal), the system changes from a phase separated steady state (Fig.~\ref{fig:1D}a) to a homogeneous steady state when $\alpha_u<-E/c$ (Fig.~\ref{fig:1D}b), and develops a limit cycle (LC) at large negative $\alpha_u$ (Fig.~\ref{fig:1D}c).  On the other hand, fixing $\alpha_u>0$ and increasing $\alpha_v$ (strain rate also feeds back to reaction) triggers nonequilibrium states with increasing complexity: LC (Fig.~\ref{fig:1D}d), period doubling (Fig.~\ref{fig:1D}e), and chaos (Fig.~\ref{fig:1D}f).  
These dynamical states caused by $\alpha_u<0$ or $\alpha_v>0$  are sustained by continuous conversion of chemical energy to mechanical energy.
While nonreciprocal couplings are known to be a generic mechanism for creating dynamic patterns in systems with conservation laws \cite{scheibner2020odd,you2020nonreciprocity,saha2020scalar,brauns2024nonreciprocal}, here we focus on their interplay with inertia.

Scaling analysis reveals the key features underlying the transitions between the qualitative behaviors exhibited in Fig.~\ref{fig:1D}. In the absence of a chemical reaction ($\chi=0$), long-wavelength deformations on the scale of the system size $L$ transport momentum by elastic waves on an inertial timescale $\tau_{\rm{el}}\sim L\sqrt{\rho/E}$, and dissipate it through viscous stresses on a diffusive (Stokes) timescale $\tau_{\eta}\sim \rho L^2/\eta$. Importantly, we note that $\tau_\eta$ depends on inertia and is distinct from the local viscoelastic relaxation time $\tau_{ve}\sim \eta/E$.
The bare reaction kinetics on the other hand proceeds on a chemical timescale $\sim\tau_c$, but nonequilibrium mechanochemical feedback can pump energy into the system at independent rates. To see this, we consider the time taken for a minimal feedback cycle (Fig.~\ref{fig:barrier}b,c). A chemically induced stress $\sim c\chi$ generates a characteristic elastic strain $|\e|\sim c\chi/E$ or viscous strain rate $|\dot{\e}|\sim c\chi/\eta$ that can in turn drive the reaction via $\tau_c\dot{\chi}\sim \alpha_u\e$ or $\tau_c\dot{\chi}\sim \alpha_v|\dot{\e}|$, completing the cycle. This gives two new time-scales, $\tau_1\sim \tau_cE/|\alpha_u c|$ and $\tau_2\sim \tau_c\eta/|\alpha_vc|$, set by the strain and strain-rate feedback terms respectively (Fig.~\ref{fig:barrier}b). When activity overwhelms the rate of mechanical dissipation, i.e., $\tau_1,\tau_2\lesssim\tau_\eta,\tau_{ve}$, we expect complex dynamics to emerge (Fig.~\ref{fig:1D}). Note that for the examples in Fig.~\ref{fig:1D} we choose $\eta=0.16$, so $\tau_\eta>\tau_{\rm{el}}$ and the system is mechanically underdamped. 

\begin{figure*}[t]
    \centering    \includegraphics[width=0.99\textwidth,keepaspectratio]{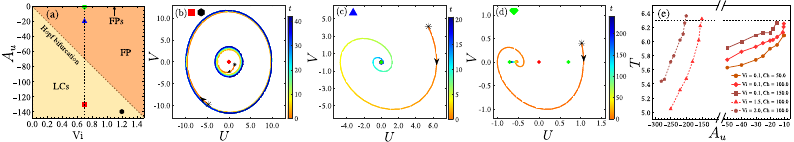}
    \caption{Dynamical phases due to nonreciprocal strain-chemistry feedback ($A_v=0, A_u<0$) in the 0D model (Eq.~\eqref{eq:0D}).   
    (a) Phase diagram ($\mathsf{Ch} = 100$).  (b-d) Typical trajectories in the $(U,V)$ plane for 
    (b) the outer (larger) LC with $\left( \mathsf{Vi}, A_u \right)=\left(0.70, -130.0\right)$ (underdamped regime) and the inner (smaller) LC with $\left( \mathsf{Vi},A_u \right)=\left(1.2,-140.0 \right)$ (overdamped regime), (c) one FP at $U_0 = X_0 = 0, V_0=0$ with $\left( \mathsf{Vi}, A_u \right)=\left(0.70, -20.0\right)$, and (d) FPs with the trajectory converging to one of them $U_+ = -X_{+} = \sqrt{1+A_u}$ with $\left( \mathsf{Vi}, A_u \right)=\left(0.70, -0.5\right)$. The FPs are shown as red (unstable) and green (stable) dots in (b-d), with the arbitrarily chosen initial conditions denoted by the black star. In panels (b-d), different symbols adjacent to the panel index represent the corresponding points in the phase diagram in (a).
    (e) Dependence of the time period $T$ on $A_u$ for different $\mathsf{Ch}$ in the LC region (in $\tau_{\rm{el}}=1$ units).
    }
    \label{fig:StrainPD}
\end{figure*}

\begin{figure*}
    \centering   \includegraphics[width=0.85\textwidth,keepaspectratio]{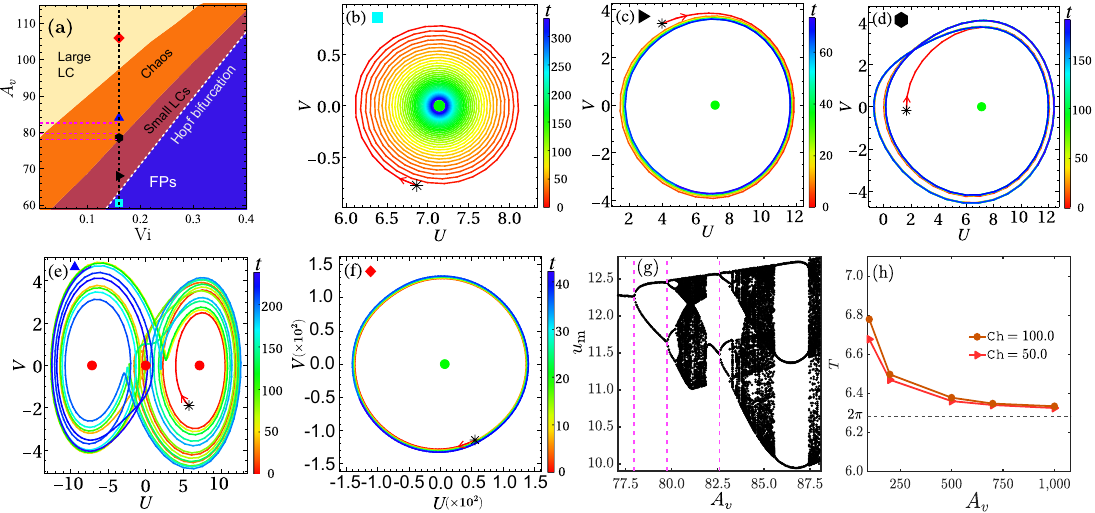}
    \caption{Dynamical phases due to nonreciprocal strain-strain rate-chemistry feedback ($A_v>0, A_u>0$) in the 0D model (Eq.~\eqref{eq:0D}). (a) Phase diagram ($\mathrm{Ch} = 100$, $A_u = 50$).  (b-f) Typical trajectories in the $(U,V)$ plane for the FPs with the trajectory converging to (b) one FP $U_+ = -X_+ = \sqrt{1+A_u}$ for $A_v=60.5$, (c) an LC around one unstable FP for $A_v=68.0$, (d) period doubling for $A_v=78.5$, (e) chaos for $A_v=84.0$, and (f) the large LC enclosing all FPs for $A_v=106.0$.  In all cases, we set $\mathsf{Vi}=0.16$ and $A_u=50.0$. The black stars in (b-f) denote the arbitrarily chosen initial condition. (g) Bifurcation diagram showing the route from LC around the FP $U_{\pm} = -X_{\pm} = \sqrt{1+A_u}$ to chaos as a function of the parameter $A_v$. Here, $u_{\rm m}$ represents the local maximum value of $U$~\cite{SM2024}. (h) Dependence of the time period $T$ on $A_v$ at $\mathsf{Vi}=0.16$ in the large LC region, approaching $2\pi$ as $A_v$ increases (in $\tau_{\rm{el}}=1$ units).
}
    \label{fig:StrainRatePD}
\end{figure*}

\noindent{\it Effective 0D model.}
We next perform a single mode Galerkin approximation to reduce the spatial complexity of the 1D model by 
projecting Eq.~\eqref{eq:1Deqs} onto the first fundamental mode of the 1D system. We set 
$u(x,t)=(1/q)U(t) \sin(qx)$, $v(x,t)=(\sqrt{E/\rho})V(t)\sin(qx)$, and $\chi(x,t)=(E/c)X(t)\cos(qx)$, with $q=2\pi/L$ to non-dimensionalize the equations and obtain 
\begin{align}\label{eq:0D}
  &\tau_{\rm{el}}\;\dot{U} = V\;,\nonumber\\
  &\tau_{\rm{el}}\;\dot{V} = - U - \mathsf{Vi}\;V - X\;,\\
  &\xm{\tau_{\rm{el}}\Ch}\;\dot{X} =  X(1-X^2) - A_u\;U- A_v\;V\;,\nonumber
\end{align}
where we have assumed $g=(4/3)(c/E)^2$ to simplify the chemical nonlinearity and set $\tau_{\rm{el}}=(L/2\pi)\sqrt{\rho/E}$ as the inertial time for a long-wavelength elastic wave.
The relative importance of viscous dissipation to inertia defines a \emph{visco-inertial} number $\Vi= 2\pi\eta/(L\sqrt{\rho E})\sim\tau_{\rm{el}}/\tau_{\eta}$. Similarly, the ratio of the chemical reaction time to that of inertia defines a \emph{chemo-inertial number} $\Ch=(2\pi\tau_c/L)\sqrt{E/\rho}=\tau_c/\tau_{\rm{el}}$. As we are primarily interested in the mechanically underdamped regime with slow chemical reactions, we expect $\tau_{\rm{el}}<\tau_\eta<\tau_c$, i.e., $\Vi<1$ and $\Ch\gg 1$. Finally, nonequilibrium activity is captured by two feedback parameters: $A_u=\alpha_u c/E\sim \tau_c/\tau_1$ and $A_v=2\pi\alpha_v c/(L\sqrt{\rho E})\sim \Vi\;\tau_c/\tau_2$. Note that when $A_u<0$, the mechanochemical coupling is nonreciprocal. With initial conditions, the four nondimensional parameters $\Vi$, $\Ch$, $A_u$, and $A_v$ completely define the response of our 0D system. Mathematically similar models appear in quantum and classical control systems  \cite{STANTON20123617,johansson2014strongly,mohammadpour2023controlling,erturk2011broadband,RevModPhys.88.035002,PhysRevB.92.115407}, metamaterials \cite{browning2019reversible} and in the physics of hearing \cite{reichenbach2014physics,martin2021mechanical}, but the characterization of the nonlinear dynamics in the context of inertial mechanochemical solids is novel to our knowledge. As in the 1D model, for steady-state, Eqs.~\eqref{eq:0D} has (i) one fixed point (FP)  $U_0 = X_0 = 0$ for $A_u<-1$, and (ii) three FPs $U_\pm = -X_\pm = \pm\sqrt{1+A_u}$ and $U_0 = X_0 = 0$ for $A_u>-1$ (at FPs, by definition, $V = 0$). 


At $A_v=0$, nonreciprocity between strain and reaction arises for $A_u<0$. Upon decreasing $A_u$, a linear stability analysis reveals first a pitchfork bifurcation where 3 FPs (bistable) transition to 1 FP (monostable) and then an oscillatory Hopf bifurcation occurs at a second threshold when $A_u\approx-\Vi\;\Ch$ \cite{matcont} (Fig.~\ref{fig:StrainPD}).
{
Figure~\ref{fig:StrainPD} shows typical trajectories in the $(U, V )$ plane for limit  cycles (b) and FPs  (c-d).}
The analytical bifurcation thresholds (derived in the SM~\cite{SM2024}) are confirmed via numerical continuation using MATCONT and shown in Fig.~\ref{fig:StrainPD}a.
This Hopf bifurcation corresponds to $\tau_1\lesssim\tau_\eta$, i.e.,
when the strain feedback actively pumps energy faster than viscous stresses can dissipate it, and it crucially depends on inertia as $\tau_\eta\sim \rho L^2/\eta$.

Near the Hopf bifurcation, the imaginary part of the unstable eigenvalue yields the time period of the LC to be $T = 2\pi \tau_{\rm{el}}/\sqrt{1-(\Vi/\Ch)}\simeq 2\pi\tau_{\rm{el}}$ for $\Ch\gg 1$ (dashed line in Fig.~\ref{fig:StrainPD}e): fast oscillations at the inertial time scale, 
distinct from previous examples of spontaneous oscillations in overdamped active solids \cite{scheibner2020non,boocock2021theory,banerjee2015propagating,shankar2024active,banerjee2019continuum}. 
{Interestingly, this instability and LC behavior persists even in the mechanically overdamped regime ($\Vi>1$, see SM~\cite{SM2024} for more details) with a period controlled by the same $\tau_{\rm{el}}$, as long as nonreciprocal feedback overwhelms the mechanical dissipation rate (Fig.~\ref{fig:StrainPD}b,e).
}

Now we consider $A_u>0, A_v>0$, where strain rate steers the  activity  (Fig.~\ref{fig:StrainRatePD}).   
By choosing $A_u>0$, we  have three FPs, with the FP at the origin ($U_0,X_0$) always  unstable. A linear stability analysis shows that upon increasing $A_v$, both nontrivial FPs $(U_\pm,X_\pm)$ lose stability via a Hopf bifurcation when $A_v\gtrsim\Ch$ (valid in the limit of slow kinetics, i.e., $\Ch\gg \Vi,A_u$, see SM~\cite{SM2024} for the full expression). This analytical bifurcation threshold (white dashed line in Fig.~\ref{fig:StrainRatePD}(a)) is confirmed by numerical analysis. In terms of physical time scales, this bifurcation condition corresponds to $\tau_2<\tau_{ve}$, i.e., the strain rate feedback time is smaller than the local viscoelastic relaxation time. 
Near onset, the small LC has a slow oscillation time period $T_{\rm{small}}\approx\pi\tau_{\rm{el}}\sqrt{2\Vi\,\Ch/(1+A_u)}\sim \sqrt{\eta\tau_c/E}$ governed by dissipative processes. Note that these timescales do not rely on inertia, and the resulting instability is smoothly connected to previously studied instabilities in overdamped active solids \cite{scheibner2020non,boocock2021theory,banerjee2015propagating,shankar2024active,banerjee2019continuum}.

\ss{Enabled by the three-dimensionality of phase space ($U,V,X$) including inertia, at large $A_v$, chaos arises}.  
We find a period doubling (Fig.~\ref{fig:StrainRatePD}d) cascade for the pair of LCs, which collide into a chaotic strange attractor (Fig.~\ref{fig:StrainRatePD}e), reminiscent of the classical Lorenz and Rossler system \cite{strogatz2024nonlinear}. Chaos develops in a large parameter regime in the $A_v-\Vi$ space (Fig.~\ref{fig:StrainRatePD}a), and is interspersed by small islands of periodic orbits (Fig.~\ref{fig:StrainRatePD}g and SM~\cite{SM2024} for Lyapunov exponent analysis). The phase boundaries separating chaos from limit cycles in Fig.~\ref{fig:StrainRatePD}(a) are obtained by analyzing phase-space trajectories obtained by direct numerical simulations. 
For even larger values of $A_v$, eventually, the system settles into a globally attracting large LC that encloses all three FPs (Fig.~\ref{fig:StrainRatePD}f).   
While the small LCs have slow oscillations, we numerically find that the large LC exhibits 
fast inertial oscillations with a time period approaching $T\approx 2\pi\tau_{\rm{el}}$ as $A_v\to\infty$ (Fig.~\ref{fig:StrainRatePD}h). 
This explains and recapitulates  the complex behavior observed in the 1D model (Fig.~\ref{fig:1D}). \nb{\ss{Whether such chaotic and fast dynamics observed for $A_v>0$ survives when $\Vi\gtrsim1$ is} left for future work.} 

\noindent{\it Conclusion.} By using nonequilibrium feedback to couple chemical reactions with the mechanics of a soft solid, we have shown how active driving can sustain autonomous deformations, fast oscillations, and complex dynamics. These results highlight the importance of inertia in active mechanochemical solids and provide a design principle for generating high quality ultrafast oscillators even in dissipative, soft solids. That feedback can also trigger chaos in active solids suggests a simple way to generate complex ``twitching'' movements, reminiscent of rudimentary behaviors in soft-bodied organisms \cite{elliott2007coordinated,armon2018ultrafast}. 
Our work paves the way to explore a new regime of chemically active solids in previously inaccessible high frequency scales, and provides a scalable and efficient platform to design autonomous soft materials from homogeneous media, without the need for complex microstructures or discretely embedded active particles.

\xm{Recent experiments suggest a route to test our predictions.
A promising platform is pH sensitive hydrogels that can achieve reversible swelling and actuation with $c/E\sim0.1-0.5$ pH$^{-1}$ (assuming $\chi\sim $pH) ~\cite{yuan_programmable,yashin2006pattern,levin2020self}. Rapid feedback can be engineered for e.g., using swelling induced mechanophore activation \cite{metze2023swelling} or strain triggered acidification \cite{ouchi2023strain} that occurs within a second, providing an estimate of $|\alpha_u|/\tau_c\sim0.1$ pH$/$s. 
When combined in a single material, this would result in $\tau_1\sim10^{-2}$ s ~\cite{SM2024}. To observe the predicted inertial Hopf bifurcation, we need $\tau_1<\tau_{\eta}$ which requires a viscosity $\eta\sim10-100$ Pa$\cdot$s (assuming  $\rho\sim 10^3~$kg/m$^3$ and $L\sim1-3~$cm), potentially achievable in low-dissipation hydrogels \cite{kim2021fracture}. 
Typical values of small molecule diffusion constants in hydrogels, $D\sim 10^{-9}$ m$^2$/s~\cite{westrin1994diffusion}, yield a diffusion time scale $L^2/D\sim 10^5$ s, much larger than  time scales of interest in this problem, justifying our neglect of molecular diffusion.  
Similar estimates can be made for  strain-rate feedback that has been qualitatively observed \textit{e.g.}, in Ref.~\cite{ouchi2023strain}. 
Biological materials, \textit{e.g.}, actomyosin or muscle tissue, can achieve strong active stresses and feedback with $c/E\sim 0.1-1$, $\tau_c\sim 10^{-3}~$s, $|\alpha_u|\sim 0.1-10$ and $|\alpha_v|/\tau_c\sim 1-10$ (using $\chi\sim$ myosin fraction) \cite{shankar2024active,bustamante2004mechanical}, leading to $\tau_1\sim 10^{-4}-0.1~$s and $\tau_2/\tau_{ve}\sim 0.1-10$. But their large viscosity $\eta\gtrsim 10^5~$Pa$\cdot$s \cite{shankar2024active,joanny2009active} suppresses the inertial instability requiring $\tau_1<\tau_\eta\lesssim 10^{-8}~$s (assuming $\rho\sim 10^3~$kg/m$^3$ and $L\sim 1~$mm), though $\tau_2<\tau_{ve}$ permits the overdamped small LC to exist \cite{banerjee2015propagating,shankar2024active,banerjee2019continuum}.
This suggests combining weak dissipation with strong feedback as a simple principle for designing inertial mechanochemical metamaterials, powered by and consisting of collectives of, \textit{e.g.}, chemically fueled microrobots \cite{yang202088}. 
}

\noindent{\it Acknowledgments.}--N.B., X.M. and B.A. acknowledge support from the US Army Research Office (Grant No. W911NF-20-2-0182). X.M., S.S. and Y.W. acknowledge support from the Office of Naval Research (MURI N00014-20-1-2479). X.M. acknowledge support from National Science Foundation
through the Materials Research Science and Engineering Center at the University of Michigan, Award No. DMR-2309029. 
This research was supported in part by grant NSF PHY-2309135 to the Kavli Institute for Theoretical Physics (KITP).


\begin{thebibliography}{81}%
\makeatletter
\providecommand \@ifxundefined [1]{%
 \@ifx{#1\undefined}
}%
\providecommand \@ifnum [1]{%
 \ifnum #1\expandafter \@firstoftwo
 \else \expandafter \@secondoftwo
 \fi
}%
\providecommand \@ifx [1]{%
 \ifx #1\expandafter \@firstoftwo
 \else \expandafter \@secondoftwo
 \fi
}%
\providecommand \natexlab [1]{#1}%
\providecommand \enquote  [1]{``#1''}%
\providecommand \bibnamefont  [1]{#1}%
\providecommand \bibfnamefont [1]{#1}%
\providecommand \citenamefont [1]{#1}%
\providecommand \href@noop [0]{\@secondoftwo}%
\providecommand \href [0]{\begingroup \@sanitize@url \@href}%
\providecommand \@href[1]{\@@startlink{#1}\@@href}%
\providecommand \@@href[1]{\endgroup#1\@@endlink}%
\providecommand \@sanitize@url [0]{\catcode `\\12\catcode `\$12\catcode
  `\&12\catcode `\#12\catcode `\^12\catcode `\_12\catcode `\%12\relax}%
\providecommand \@@startlink[1]{}%
\providecommand \@@endlink[0]{}%
\providecommand \url  [0]{\begingroup\@sanitize@url \@url }%
\providecommand \@url [1]{\endgroup\@href {#1}{\urlprefix }}%
\providecommand \urlprefix  [0]{URL }%
\providecommand \Eprint [0]{\href }%
\providecommand \doibase [0]{http://dx.doi.org/}%
\providecommand \selectlanguage [0]{\@gobble}%
\providecommand \bibinfo  [0]{\@secondoftwo}%
\providecommand \bibfield  [0]{\@secondoftwo}%
\providecommand \translation [1]{[#1]}%
\providecommand \BibitemOpen [0]{}%
\providecommand \bibitemStop [0]{}%
\providecommand \bibitemNoStop [0]{.\EOS\space}%
\providecommand \EOS [0]{\spacefactor3000\relax}%
\providecommand \BibitemShut  [1]{\csname bibitem#1\endcsname}%
\let\auto@bib@innerbib\@empty
\bibitem [{\citenamefont {Asaka}\ and\ \citenamefont
  {Okuzaki}(2019)}]{asaka2019soft}%
  \BibitemOpen
  \bibfield  {author} {\bibinfo {author} {\bibfnamefont {K.}~\bibnamefont
  {Asaka}}\ and\ \bibinfo {author} {\bibfnamefont {H.}~\bibnamefont
  {Okuzaki}},\ }\href@noop {} {\emph {\bibinfo {title} {Soft actuators:
  materials, modeling, applications, and future perspectives}}}\ (\bibinfo
  {publisher} {Springer Nature},\ \bibinfo {year} {2019})\BibitemShut {NoStop}%
\bibitem [{\citenamefont {Rus}\ and\ \citenamefont
  {Tolley}(2015)}]{rus2015design}%
  \BibitemOpen
  \bibfield  {author} {\bibinfo {author} {\bibfnamefont {D.}~\bibnamefont
  {Rus}}\ and\ \bibinfo {author} {\bibfnamefont {M.~T.}\ \bibnamefont
  {Tolley}},\ }\href@noop {} {\bibfield  {journal} {\bibinfo  {journal}
  {Nature}\ }\textbf {\bibinfo {volume} {521}},\ \bibinfo {pages} {467}
  (\bibinfo {year} {2015})}\BibitemShut {NoStop}%
\bibitem [{\citenamefont {Li}\ \emph {et~al.}(2022)\citenamefont {Li},
  \citenamefont {Pal}, \citenamefont {Aghakhani}, \citenamefont
  {Pena-Francesch},\ and\ \citenamefont {Sitti}}]{li2022soft}%
  \BibitemOpen
  \bibfield  {author} {\bibinfo {author} {\bibfnamefont {M.}~\bibnamefont
  {Li}}, \bibinfo {author} {\bibfnamefont {A.}~\bibnamefont {Pal}}, \bibinfo
  {author} {\bibfnamefont {A.}~\bibnamefont {Aghakhani}}, \bibinfo {author}
  {\bibfnamefont {A.}~\bibnamefont {Pena-Francesch}}, \ and\ \bibinfo {author}
  {\bibfnamefont {M.}~\bibnamefont {Sitti}},\ }\href@noop {} {\bibfield
  {journal} {\bibinfo  {journal} {Nature Reviews Materials}\ }\textbf {\bibinfo
  {volume} {7}},\ \bibinfo {pages} {235} (\bibinfo {year} {2022})}\BibitemShut
  {NoStop}%
\bibitem [{\citenamefont {de~Gennes}(1997)}]{de1997semi}%
  \BibitemOpen
  \bibfield  {author} {\bibinfo {author} {\bibfnamefont {P.-G.}\ \bibnamefont
  {de~Gennes}},\ }\href@noop {} {\bibfield  {journal} {\bibinfo  {journal}
  {Comptes Rendus de l'Academie des Sciences Series IIB Mechanics Physics
  Chemistry Astronomy}\ }\textbf {\bibinfo {volume} {5}},\ \bibinfo {pages}
  {343} (\bibinfo {year} {1997})}\BibitemShut {NoStop}%
\bibitem [{\citenamefont {Mirvakili}\ and\ \citenamefont
  {Hunter}(2018)}]{mirvakili2018artificial}%
  \BibitemOpen
  \bibfield  {author} {\bibinfo {author} {\bibfnamefont {S.~M.}\ \bibnamefont
  {Mirvakili}}\ and\ \bibinfo {author} {\bibfnamefont {I.~W.}\ \bibnamefont
  {Hunter}},\ }\href@noop {} {\bibfield  {journal} {\bibinfo  {journal}
  {Advanced Materials}\ }\textbf {\bibinfo {volume} {30}},\ \bibinfo {pages}
  {1704407} (\bibinfo {year} {2018})}\BibitemShut {NoStop}%
\bibitem [{\citenamefont {He}\ \emph {et~al.}(2021)\citenamefont {He},
  \citenamefont {Wang}, \citenamefont {Wang}, \citenamefont {Wang},
  \citenamefont {Li}, \citenamefont {Annapooranan}, \citenamefont {Zeng},
  \citenamefont {Chen},\ and\ \citenamefont {Cai}}]{he2021electrospun}%
  \BibitemOpen
  \bibfield  {author} {\bibinfo {author} {\bibfnamefont {Q.}~\bibnamefont
  {He}}, \bibinfo {author} {\bibfnamefont {Z.}~\bibnamefont {Wang}}, \bibinfo
  {author} {\bibfnamefont {Y.}~\bibnamefont {Wang}}, \bibinfo {author}
  {\bibfnamefont {Z.}~\bibnamefont {Wang}}, \bibinfo {author} {\bibfnamefont
  {C.}~\bibnamefont {Li}}, \bibinfo {author} {\bibfnamefont {R.}~\bibnamefont
  {Annapooranan}}, \bibinfo {author} {\bibfnamefont {J.}~\bibnamefont {Zeng}},
  \bibinfo {author} {\bibfnamefont {R.}~\bibnamefont {Chen}}, \ and\ \bibinfo
  {author} {\bibfnamefont {S.}~\bibnamefont {Cai}},\ }\href@noop {} {\bibfield
  {journal} {\bibinfo  {journal} {Science robotics}\ }\textbf {\bibinfo
  {volume} {6}},\ \bibinfo {pages} {eabi9704} (\bibinfo {year}
  {2021})}\BibitemShut {NoStop}%
\bibitem [{\citenamefont {Bas}\ \emph {et~al.}(2021)\citenamefont {Bas},
  \citenamefont {Gorissen}, \citenamefont {Luposchainsky}, \citenamefont
  {Shabab}, \citenamefont {Bertoldi},\ and\ \citenamefont
  {Hutmacher}}]{bas2021ultrafast}%
  \BibitemOpen
  \bibfield  {author} {\bibinfo {author} {\bibfnamefont {O.}~\bibnamefont
  {Bas}}, \bibinfo {author} {\bibfnamefont {B.}~\bibnamefont {Gorissen}},
  \bibinfo {author} {\bibfnamefont {S.}~\bibnamefont {Luposchainsky}}, \bibinfo
  {author} {\bibfnamefont {T.}~\bibnamefont {Shabab}}, \bibinfo {author}
  {\bibfnamefont {K.}~\bibnamefont {Bertoldi}}, \ and\ \bibinfo {author}
  {\bibfnamefont {D.~W.}\ \bibnamefont {Hutmacher}},\ }\href@noop {} {\bibfield
   {journal} {\bibinfo  {journal} {Multifunctional Materials}\ }\textbf
  {\bibinfo {volume} {4}},\ \bibinfo {pages} {045001} (\bibinfo {year}
  {2021})}\BibitemShut {NoStop}%
\bibitem [{\citenamefont {Skotheim}\ and\ \citenamefont
  {Mahadevan}(2005)}]{skotheim2005physical}%
  \BibitemOpen
  \bibfield  {author} {\bibinfo {author} {\bibfnamefont {J.~M.}\ \bibnamefont
  {Skotheim}}\ and\ \bibinfo {author} {\bibfnamefont {L.}~\bibnamefont
  {Mahadevan}},\ }\href@noop {} {\bibfield  {journal} {\bibinfo  {journal}
  {Science}\ }\textbf {\bibinfo {volume} {308}},\ \bibinfo {pages} {1308}
  (\bibinfo {year} {2005})}\BibitemShut {NoStop}%
\bibitem [{\citenamefont {Lee}\ \emph {et~al.}(2010)\citenamefont {Lee},
  \citenamefont {Xia},\ and\ \citenamefont {Fang}}]{lee2010first}%
  \BibitemOpen
  \bibfield  {author} {\bibinfo {author} {\bibfnamefont {H.}~\bibnamefont
  {Lee}}, \bibinfo {author} {\bibfnamefont {C.}~\bibnamefont {Xia}}, \ and\
  \bibinfo {author} {\bibfnamefont {N.~X.}\ \bibnamefont {Fang}},\ }\href@noop
  {} {\bibfield  {journal} {\bibinfo  {journal} {Soft Matter}\ }\textbf
  {\bibinfo {volume} {6}},\ \bibinfo {pages} {4342} (\bibinfo {year}
  {2010})}\BibitemShut {NoStop}%
\bibitem [{\citenamefont {Gorissen}\ \emph {et~al.}(2020)\citenamefont
  {Gorissen}, \citenamefont {Melancon}, \citenamefont {Vasios}, \citenamefont
  {Torbati},\ and\ \citenamefont {Bertoldi}}]{gorissen2020inflatable}%
  \BibitemOpen
  \bibfield  {author} {\bibinfo {author} {\bibfnamefont {B.}~\bibnamefont
  {Gorissen}}, \bibinfo {author} {\bibfnamefont {D.}~\bibnamefont {Melancon}},
  \bibinfo {author} {\bibfnamefont {N.}~\bibnamefont {Vasios}}, \bibinfo
  {author} {\bibfnamefont {M.}~\bibnamefont {Torbati}}, \ and\ \bibinfo
  {author} {\bibfnamefont {K.}~\bibnamefont {Bertoldi}},\ }\href@noop {}
  {\bibfield  {journal} {\bibinfo  {journal} {Science Robotics}\ }\textbf
  {\bibinfo {volume} {5}},\ \bibinfo {pages} {eabb1967} (\bibinfo {year}
  {2020})}\BibitemShut {NoStop}%
\bibitem [{\citenamefont {Chen}\ \emph {et~al.}(2019)\citenamefont {Chen},
  \citenamefont {Zhao}, \citenamefont {Mao}, \citenamefont {Chirarattananon},
  \citenamefont {Helbling}, \citenamefont {Hyun}, \citenamefont {Clarke},\ and\
  \citenamefont {Wood}}]{chen2019controlled}%
  \BibitemOpen
  \bibfield  {author} {\bibinfo {author} {\bibfnamefont {Y.}~\bibnamefont
  {Chen}}, \bibinfo {author} {\bibfnamefont {H.}~\bibnamefont {Zhao}}, \bibinfo
  {author} {\bibfnamefont {J.}~\bibnamefont {Mao}}, \bibinfo {author}
  {\bibfnamefont {P.}~\bibnamefont {Chirarattananon}}, \bibinfo {author}
  {\bibfnamefont {E.~F.}\ \bibnamefont {Helbling}}, \bibinfo {author}
  {\bibfnamefont {N.-s.~P.}\ \bibnamefont {Hyun}}, \bibinfo {author}
  {\bibfnamefont {D.~R.}\ \bibnamefont {Clarke}}, \ and\ \bibinfo {author}
  {\bibfnamefont {R.~J.}\ \bibnamefont {Wood}},\ }\href@noop {} {\bibfield
  {journal} {\bibinfo  {journal} {Nature}\ }\textbf {\bibinfo {volume} {575}},\
  \bibinfo {pages} {324} (\bibinfo {year} {2019})}\BibitemShut {NoStop}%
\bibitem [{\citenamefont {Mao}\ \emph {et~al.}(2022)\citenamefont {Mao},
  \citenamefont {Schiller}, \citenamefont {Danninger}, \citenamefont
  {Hailegnaw}, \citenamefont {Hartmann}, \citenamefont {Stockinger},
  \citenamefont {Drack}, \citenamefont {Arnold},\ and\ \citenamefont
  {Kaltenbrunner}}]{mao2022ultrafast}%
  \BibitemOpen
  \bibfield  {author} {\bibinfo {author} {\bibfnamefont {G.}~\bibnamefont
  {Mao}}, \bibinfo {author} {\bibfnamefont {D.}~\bibnamefont {Schiller}},
  \bibinfo {author} {\bibfnamefont {D.}~\bibnamefont {Danninger}}, \bibinfo
  {author} {\bibfnamefont {B.}~\bibnamefont {Hailegnaw}}, \bibinfo {author}
  {\bibfnamefont {F.}~\bibnamefont {Hartmann}}, \bibinfo {author}
  {\bibfnamefont {T.}~\bibnamefont {Stockinger}}, \bibinfo {author}
  {\bibfnamefont {M.}~\bibnamefont {Drack}}, \bibinfo {author} {\bibfnamefont
  {N.}~\bibnamefont {Arnold}}, \ and\ \bibinfo {author} {\bibfnamefont
  {M.}~\bibnamefont {Kaltenbrunner}},\ }\href@noop {} {\bibfield  {journal}
  {\bibinfo  {journal} {Nature communications}\ }\textbf {\bibinfo {volume}
  {13}},\ \bibinfo {pages} {4456} (\bibinfo {year} {2022})}\BibitemShut
  {NoStop}%
\bibitem [{\citenamefont {Shepherd}\ \emph {et~al.}(2013)\citenamefont
  {Shepherd}, \citenamefont {Stokes}, \citenamefont {Freake}, \citenamefont
  {Barber}, \citenamefont {Snyder}, \citenamefont {Mazzeo}, \citenamefont
  {Cademartiri}, \citenamefont {Morin},\ and\ \citenamefont
  {Whitesides}}]{shepherd2013using}%
  \BibitemOpen
  \bibfield  {author} {\bibinfo {author} {\bibfnamefont {R.~F.}\ \bibnamefont
  {Shepherd}}, \bibinfo {author} {\bibfnamefont {A.~A.}\ \bibnamefont
  {Stokes}}, \bibinfo {author} {\bibfnamefont {J.}~\bibnamefont {Freake}},
  \bibinfo {author} {\bibfnamefont {J.}~\bibnamefont {Barber}}, \bibinfo
  {author} {\bibfnamefont {P.~W.}\ \bibnamefont {Snyder}}, \bibinfo {author}
  {\bibfnamefont {A.~D.}\ \bibnamefont {Mazzeo}}, \bibinfo {author}
  {\bibfnamefont {L.}~\bibnamefont {Cademartiri}}, \bibinfo {author}
  {\bibfnamefont {S.~A.}\ \bibnamefont {Morin}}, \ and\ \bibinfo {author}
  {\bibfnamefont {G.~M.}\ \bibnamefont {Whitesides}},\ }\href@noop {}
  {\bibfield  {journal} {\bibinfo  {journal} {Angewandte Chemie International
  Edition}\ }\textbf {\bibinfo {volume} {52}},\ \bibinfo {pages} {2892}
  (\bibinfo {year} {2013})}\BibitemShut {NoStop}%
\bibitem [{\citenamefont {Aubin}\ \emph {et~al.}(2023)\citenamefont {Aubin},
  \citenamefont {Heisser}, \citenamefont {Peretz}, \citenamefont {Timko},
  \citenamefont {Lo}, \citenamefont {Helbling}, \citenamefont {Sobhani},
  \citenamefont {Gat},\ and\ \citenamefont {Shepherd}}]{aubin2023powerful}%
  \BibitemOpen
  \bibfield  {author} {\bibinfo {author} {\bibfnamefont {C.~A.}\ \bibnamefont
  {Aubin}}, \bibinfo {author} {\bibfnamefont {R.~H.}\ \bibnamefont {Heisser}},
  \bibinfo {author} {\bibfnamefont {O.}~\bibnamefont {Peretz}}, \bibinfo
  {author} {\bibfnamefont {J.}~\bibnamefont {Timko}}, \bibinfo {author}
  {\bibfnamefont {J.}~\bibnamefont {Lo}}, \bibinfo {author} {\bibfnamefont
  {E.~F.}\ \bibnamefont {Helbling}}, \bibinfo {author} {\bibfnamefont
  {S.}~\bibnamefont {Sobhani}}, \bibinfo {author} {\bibfnamefont {A.~D.}\
  \bibnamefont {Gat}}, \ and\ \bibinfo {author} {\bibfnamefont {R.~F.}\
  \bibnamefont {Shepherd}},\ }\href@noop {} {\bibfield  {journal} {\bibinfo
  {journal} {Science}\ }\textbf {\bibinfo {volume} {381}},\ \bibinfo {pages}
  {1212} (\bibinfo {year} {2023})}\BibitemShut {NoStop}%
\bibitem [{\citenamefont {Armon}\ \emph {et~al.}(2018)\citenamefont {Armon},
  \citenamefont {Bull}, \citenamefont {Aranda-Diaz},\ and\ \citenamefont
  {Prakash}}]{armon2018ultrafast}%
  \BibitemOpen
  \bibfield  {author} {\bibinfo {author} {\bibfnamefont {S.}~\bibnamefont
  {Armon}}, \bibinfo {author} {\bibfnamefont {M.~S.}\ \bibnamefont {Bull}},
  \bibinfo {author} {\bibfnamefont {A.}~\bibnamefont {Aranda-Diaz}}, \ and\
  \bibinfo {author} {\bibfnamefont {M.}~\bibnamefont {Prakash}},\ }\href@noop
  {} {\bibfield  {journal} {\bibinfo  {journal} {Proceedings of the National
  Academy of Sciences}\ }\textbf {\bibinfo {volume} {115}},\ \bibinfo {pages}
  {E10333} (\bibinfo {year} {2018})}\BibitemShut {NoStop}%
\bibitem [{\citenamefont {Shankar}\ and\ \citenamefont
  {Mahadevan}(2024)}]{shankar2024active}%
  \BibitemOpen
  \bibfield  {author} {\bibinfo {author} {\bibfnamefont {S.}~\bibnamefont
  {Shankar}}\ and\ \bibinfo {author} {\bibfnamefont {L.}~\bibnamefont
  {Mahadevan}},\ }\href@noop {} {\bibfield  {journal} {\bibinfo  {journal}
  {Nature Physics}\ }\textbf {\bibinfo {volume} {20}},\ \bibinfo {pages} {1501}
  (\bibinfo {year} {2024})}\BibitemShut {NoStop}%
\bibitem [{\citenamefont {Banerjee}\ and\ \citenamefont
  {Marchetti}(2019)}]{banerjee2019continuum}%
  \BibitemOpen
  \bibfield  {author} {\bibinfo {author} {\bibfnamefont {S.}~\bibnamefont
  {Banerjee}}\ and\ \bibinfo {author} {\bibfnamefont {M.~C.}\ \bibnamefont
  {Marchetti}},\ }\href@noop {} {\bibfield  {journal} {\bibinfo  {journal}
  {Cell migrations: Causes and functions}\ ,\ \bibinfo {pages} {45}} (\bibinfo
  {year} {2019})}\BibitemShut {NoStop}%
\bibitem [{\citenamefont {Noll}\ \emph {et~al.}(2017)\citenamefont {Noll},
  \citenamefont {Mani}, \citenamefont {Heemskerk}, \citenamefont {Streichan},\
  and\ \citenamefont {Shraiman}}]{noll2017active}%
  \BibitemOpen
  \bibfield  {author} {\bibinfo {author} {\bibfnamefont {N.}~\bibnamefont
  {Noll}}, \bibinfo {author} {\bibfnamefont {M.}~\bibnamefont {Mani}}, \bibinfo
  {author} {\bibfnamefont {I.}~\bibnamefont {Heemskerk}}, \bibinfo {author}
  {\bibfnamefont {S.~J.}\ \bibnamefont {Streichan}}, \ and\ \bibinfo {author}
  {\bibfnamefont {B.~I.}\ \bibnamefont {Shraiman}},\ }\href@noop {} {\bibfield
  {journal} {\bibinfo  {journal} {Nature physics}\ }\textbf {\bibinfo {volume}
  {13}},\ \bibinfo {pages} {1221} (\bibinfo {year} {2017})}\BibitemShut
  {NoStop}%
\bibitem [{\citenamefont {Boocock}\ \emph {et~al.}(2021)\citenamefont
  {Boocock}, \citenamefont {Hino}, \citenamefont {Ruzickova}, \citenamefont
  {Hirashima},\ and\ \citenamefont {Hannezo}}]{boocock2021theory}%
  \BibitemOpen
  \bibfield  {author} {\bibinfo {author} {\bibfnamefont {D.}~\bibnamefont
  {Boocock}}, \bibinfo {author} {\bibfnamefont {N.}~\bibnamefont {Hino}},
  \bibinfo {author} {\bibfnamefont {N.}~\bibnamefont {Ruzickova}}, \bibinfo
  {author} {\bibfnamefont {T.}~\bibnamefont {Hirashima}}, \ and\ \bibinfo
  {author} {\bibfnamefont {E.}~\bibnamefont {Hannezo}},\ }\href@noop {}
  {\bibfield  {journal} {\bibinfo  {journal} {Nature physics}\ }\textbf
  {\bibinfo {volume} {17}},\ \bibinfo {pages} {267} (\bibinfo {year}
  {2021})}\BibitemShut {NoStop}%
\bibitem [{\citenamefont {Tan}\ \emph {et~al.}(2022)\citenamefont {Tan},
  \citenamefont {Mietke}, \citenamefont {Li}, \citenamefont {Chen},
  \citenamefont {Higinbotham}, \citenamefont {Foster}, \citenamefont {Gokhale},
  \citenamefont {Dunkel},\ and\ \citenamefont {Fakhri}}]{tan2022odd}%
  \BibitemOpen
  \bibfield  {author} {\bibinfo {author} {\bibfnamefont {T.~H.}\ \bibnamefont
  {Tan}}, \bibinfo {author} {\bibfnamefont {A.}~\bibnamefont {Mietke}},
  \bibinfo {author} {\bibfnamefont {J.}~\bibnamefont {Li}}, \bibinfo {author}
  {\bibfnamefont {Y.}~\bibnamefont {Chen}}, \bibinfo {author} {\bibfnamefont
  {H.}~\bibnamefont {Higinbotham}}, \bibinfo {author} {\bibfnamefont {P.~J.}\
  \bibnamefont {Foster}}, \bibinfo {author} {\bibfnamefont {S.}~\bibnamefont
  {Gokhale}}, \bibinfo {author} {\bibfnamefont {J.}~\bibnamefont {Dunkel}}, \
  and\ \bibinfo {author} {\bibfnamefont {N.}~\bibnamefont {Fakhri}},\
  }\href@noop {} {\bibfield  {journal} {\bibinfo  {journal} {Nature}\ }\textbf
  {\bibinfo {volume} {607}},\ \bibinfo {pages} {287} (\bibinfo {year}
  {2022})}\BibitemShut {NoStop}%
\bibitem [{\citenamefont {Xu}\ \emph {et~al.}(2023)\citenamefont {Xu},
  \citenamefont {Huang}, \citenamefont {Zhang},\ and\ \citenamefont
  {Wu}}]{xu2023autonomous}%
  \BibitemOpen
  \bibfield  {author} {\bibinfo {author} {\bibfnamefont {H.}~\bibnamefont
  {Xu}}, \bibinfo {author} {\bibfnamefont {Y.}~\bibnamefont {Huang}}, \bibinfo
  {author} {\bibfnamefont {R.}~\bibnamefont {Zhang}}, \ and\ \bibinfo {author}
  {\bibfnamefont {Y.}~\bibnamefont {Wu}},\ }\href@noop {} {\bibfield  {journal}
  {\bibinfo  {journal} {Nature Physics}\ }\textbf {\bibinfo {volume} {19}},\
  \bibinfo {pages} {46} (\bibinfo {year} {2023})}\BibitemShut {NoStop}%
\bibitem [{\citenamefont {Yashin}\ and\ \citenamefont
  {Balazs}(2006)}]{yashin2006pattern}%
  \BibitemOpen
  \bibfield  {author} {\bibinfo {author} {\bibfnamefont {V.~V.}\ \bibnamefont
  {Yashin}}\ and\ \bibinfo {author} {\bibfnamefont {A.~C.}\ \bibnamefont
  {Balazs}},\ }\href@noop {} {\bibfield  {journal} {\bibinfo  {journal}
  {Science}\ }\textbf {\bibinfo {volume} {314}},\ \bibinfo {pages} {798}
  (\bibinfo {year} {2006})}\BibitemShut {NoStop}%
\bibitem [{\citenamefont {Yashin}\ \emph {et~al.}(2012)\citenamefont {Yashin},
  \citenamefont {Kuksenok}, \citenamefont {Dayal},\ and\ \citenamefont
  {Balazs}}]{yashin2012mechano}%
  \BibitemOpen
  \bibfield  {author} {\bibinfo {author} {\bibfnamefont {V.~V.}\ \bibnamefont
  {Yashin}}, \bibinfo {author} {\bibfnamefont {O.}~\bibnamefont {Kuksenok}},
  \bibinfo {author} {\bibfnamefont {P.}~\bibnamefont {Dayal}}, \ and\ \bibinfo
  {author} {\bibfnamefont {A.~C.}\ \bibnamefont {Balazs}},\ }\href@noop {}
  {\bibfield  {journal} {\bibinfo  {journal} {Reports on progress in physics}\
  }\textbf {\bibinfo {volume} {75}},\ \bibinfo {pages} {066601} (\bibinfo
  {year} {2012})}\BibitemShut {NoStop}%
\bibitem [{\citenamefont {Levin}\ \emph {et~al.}(2020)\citenamefont {Levin},
  \citenamefont {Deegan},\ and\ \citenamefont {Sharon}}]{levin2020self}%
  \BibitemOpen
  \bibfield  {author} {\bibinfo {author} {\bibfnamefont {I.}~\bibnamefont
  {Levin}}, \bibinfo {author} {\bibfnamefont {R.}~\bibnamefont {Deegan}}, \
  and\ \bibinfo {author} {\bibfnamefont {E.}~\bibnamefont {Sharon}},\
  }\href@noop {} {\bibfield  {journal} {\bibinfo  {journal} {Physical Review
  Letters}\ }\textbf {\bibinfo {volume} {125}},\ \bibinfo {pages} {178001}
  (\bibinfo {year} {2020})}\BibitemShut {NoStop}%
\bibitem [{\citenamefont {Baconnier}\ \emph {et~al.}(2022)\citenamefont
  {Baconnier}, \citenamefont {Shohat}, \citenamefont {L{\'o}pez}, \citenamefont
  {Coulais}, \citenamefont {D{\'e}mery}, \citenamefont {D{\"u}ring},\ and\
  \citenamefont {Dauchot}}]{baconnier2022selective}%
  \BibitemOpen
  \bibfield  {author} {\bibinfo {author} {\bibfnamefont {P.}~\bibnamefont
  {Baconnier}}, \bibinfo {author} {\bibfnamefont {D.}~\bibnamefont {Shohat}},
  \bibinfo {author} {\bibfnamefont {C.~H.}\ \bibnamefont {L{\'o}pez}}, \bibinfo
  {author} {\bibfnamefont {C.}~\bibnamefont {Coulais}}, \bibinfo {author}
  {\bibfnamefont {V.}~\bibnamefont {D{\'e}mery}}, \bibinfo {author}
  {\bibfnamefont {G.}~\bibnamefont {D{\"u}ring}}, \ and\ \bibinfo {author}
  {\bibfnamefont {O.}~\bibnamefont {Dauchot}},\ }\href@noop {} {\bibfield
  {journal} {\bibinfo  {journal} {Nature Physics}\ }\textbf {\bibinfo {volume}
  {18}},\ \bibinfo {pages} {1234} (\bibinfo {year} {2022})}\BibitemShut
  {NoStop}%
\bibitem [{\citenamefont {Scheibner}\ \emph
  {et~al.}(2020{\natexlab{a}})\citenamefont {Scheibner}, \citenamefont
  {Souslov}, \citenamefont {Banerjee}, \citenamefont {Sur{\'o}wka},
  \citenamefont {Irvine},\ and\ \citenamefont {Vitelli}}]{scheibner2020odd}%
  \BibitemOpen
  \bibfield  {author} {\bibinfo {author} {\bibfnamefont {C.}~\bibnamefont
  {Scheibner}}, \bibinfo {author} {\bibfnamefont {A.}~\bibnamefont {Souslov}},
  \bibinfo {author} {\bibfnamefont {D.}~\bibnamefont {Banerjee}}, \bibinfo
  {author} {\bibfnamefont {P.}~\bibnamefont {Sur{\'o}wka}}, \bibinfo {author}
  {\bibfnamefont {W.~T.}\ \bibnamefont {Irvine}}, \ and\ \bibinfo {author}
  {\bibfnamefont {V.}~\bibnamefont {Vitelli}},\ }\href@noop {} {\bibfield
  {journal} {\bibinfo  {journal} {Nature Physics}\ }\textbf {\bibinfo {volume}
  {16}},\ \bibinfo {pages} {475} (\bibinfo {year}
  {2020}{\natexlab{a}})}\BibitemShut {NoStop}%
\bibitem [{\citenamefont {Banerjee}\ \emph {et~al.}(2015)\citenamefont
  {Banerjee}, \citenamefont {Utuje},\ and\ \citenamefont
  {Marchetti}}]{banerjee2015propagating}%
  \BibitemOpen
  \bibfield  {author} {\bibinfo {author} {\bibfnamefont {S.}~\bibnamefont
  {Banerjee}}, \bibinfo {author} {\bibfnamefont {K.~J.}\ \bibnamefont {Utuje}},
  \ and\ \bibinfo {author} {\bibfnamefont {M.~C.}\ \bibnamefont {Marchetti}},\
  }\href@noop {} {\bibfield  {journal} {\bibinfo  {journal} {Physical review
  letters}\ }\textbf {\bibinfo {volume} {114}},\ \bibinfo {pages} {228101}
  (\bibinfo {year} {2015})}\BibitemShut {NoStop}%
\bibitem [{\citenamefont {Maitra}\ and\ \citenamefont
  {Ramaswamy}(2019)}]{maitra2019oriented}%
  \BibitemOpen
  \bibfield  {author} {\bibinfo {author} {\bibfnamefont {A.}~\bibnamefont
  {Maitra}}\ and\ \bibinfo {author} {\bibfnamefont {S.}~\bibnamefont
  {Ramaswamy}},\ }\href@noop {} {\bibfield  {journal} {\bibinfo  {journal}
  {Physical review letters}\ }\textbf {\bibinfo {volume} {123}},\ \bibinfo
  {pages} {238001} (\bibinfo {year} {2019})}\BibitemShut {NoStop}%
\bibitem [{\citenamefont {Fruchart}\ \emph {et~al.}(2023)\citenamefont
  {Fruchart}, \citenamefont {Scheibner},\ and\ \citenamefont
  {Vitelli}}]{fruchart2023odd}%
  \BibitemOpen
  \bibfield  {author} {\bibinfo {author} {\bibfnamefont {M.}~\bibnamefont
  {Fruchart}}, \bibinfo {author} {\bibfnamefont {C.}~\bibnamefont {Scheibner}},
  \ and\ \bibinfo {author} {\bibfnamefont {V.}~\bibnamefont {Vitelli}},\
  }\href@noop {} {\bibfield  {journal} {\bibinfo  {journal} {Annual Review of
  Condensed Matter Physics}\ }\textbf {\bibinfo {volume} {14}},\ \bibinfo
  {pages} {471} (\bibinfo {year} {2023})}\BibitemShut {NoStop}%
\bibitem [{\citenamefont {Notbohm}\ \emph {et~al.}(2016)\citenamefont
  {Notbohm}, \citenamefont {Banerjee}, \citenamefont {Utuje}, \citenamefont
  {Gweon}, \citenamefont {Jang}, \citenamefont {Park}, \citenamefont {Shin},
  \citenamefont {Butler}, \citenamefont {Fredberg},\ and\ \citenamefont
  {Marchetti}}]{notbohm2016cellular}%
  \BibitemOpen
  \bibfield  {author} {\bibinfo {author} {\bibfnamefont {J.}~\bibnamefont
  {Notbohm}}, \bibinfo {author} {\bibfnamefont {S.}~\bibnamefont {Banerjee}},
  \bibinfo {author} {\bibfnamefont {K.~J.}\ \bibnamefont {Utuje}}, \bibinfo
  {author} {\bibfnamefont {B.}~\bibnamefont {Gweon}}, \bibinfo {author}
  {\bibfnamefont {H.}~\bibnamefont {Jang}}, \bibinfo {author} {\bibfnamefont
  {Y.}~\bibnamefont {Park}}, \bibinfo {author} {\bibfnamefont {J.}~\bibnamefont
  {Shin}}, \bibinfo {author} {\bibfnamefont {J.~P.}\ \bibnamefont {Butler}},
  \bibinfo {author} {\bibfnamefont {J.~J.}\ \bibnamefont {Fredberg}}, \ and\
  \bibinfo {author} {\bibfnamefont {M.~C.}\ \bibnamefont {Marchetti}},\
  }\href@noop {} {\bibfield  {journal} {\bibinfo  {journal} {Biophysical
  journal}\ }\textbf {\bibinfo {volume} {110}},\ \bibinfo {pages} {2729}
  (\bibinfo {year} {2016})}\BibitemShut {NoStop}%
\bibitem [{\citenamefont {Banerjee}\ \emph {et~al.}(2017)\citenamefont
  {Banerjee}, \citenamefont {Munjal}, \citenamefont {Lecuit},\ and\
  \citenamefont {Rao}}]{banerjee2017actomyosin}%
  \BibitemOpen
  \bibfield  {author} {\bibinfo {author} {\bibfnamefont {D.~S.}\ \bibnamefont
  {Banerjee}}, \bibinfo {author} {\bibfnamefont {A.}~\bibnamefont {Munjal}},
  \bibinfo {author} {\bibfnamefont {T.}~\bibnamefont {Lecuit}}, \ and\ \bibinfo
  {author} {\bibfnamefont {M.}~\bibnamefont {Rao}},\ }\href@noop {} {\bibfield
  {journal} {\bibinfo  {journal} {Nature communications}\ }\textbf {\bibinfo
  {volume} {8}},\ \bibinfo {pages} {1121} (\bibinfo {year} {2017})}\BibitemShut
  {NoStop}%
\bibitem [{\citenamefont {Parmar}\ \emph {et~al.}(2025)\citenamefont {Parmar},
  \citenamefont {Dow}, \citenamefont {Pruitt},\ and\ \citenamefont
  {Marchetti}}]{parmar2025spontaneous}%
  \BibitemOpen
  \bibfield  {author} {\bibinfo {author} {\bibfnamefont {T.}~\bibnamefont
  {Parmar}}, \bibinfo {author} {\bibfnamefont {L.~P.}\ \bibnamefont {Dow}},
  \bibinfo {author} {\bibfnamefont {B.~L.}\ \bibnamefont {Pruitt}}, \ and\
  \bibinfo {author} {\bibfnamefont {M.~C.}\ \bibnamefont {Marchetti}},\
  }\href@noop {} {\bibfield  {journal} {\bibinfo  {journal} {PRX Life}\
  }\textbf {\bibinfo {volume} {3}},\ \bibinfo {pages} {013002} (\bibinfo {year}
  {2025})}\BibitemShut {NoStop}%
\bibitem [{\citenamefont {Chao}\ \emph {et~al.}(2024)\citenamefont {Chao},
  \citenamefont {Gokhale}, \citenamefont {Lin}, \citenamefont {Hastewell},
  \citenamefont {Bacanu}, \citenamefont {Chen}, \citenamefont {Li},
  \citenamefont {Liu}, \citenamefont {Lee}, \citenamefont {Dunkel} \emph
  {et~al.}}]{chao2024selective}%
  \BibitemOpen
  \bibfield  {author} {\bibinfo {author} {\bibfnamefont {Y.-C.}\ \bibnamefont
  {Chao}}, \bibinfo {author} {\bibfnamefont {S.}~\bibnamefont {Gokhale}},
  \bibinfo {author} {\bibfnamefont {L.}~\bibnamefont {Lin}}, \bibinfo {author}
  {\bibfnamefont {A.}~\bibnamefont {Hastewell}}, \bibinfo {author}
  {\bibfnamefont {A.}~\bibnamefont {Bacanu}}, \bibinfo {author} {\bibfnamefont
  {Y.}~\bibnamefont {Chen}}, \bibinfo {author} {\bibfnamefont {J.}~\bibnamefont
  {Li}}, \bibinfo {author} {\bibfnamefont {J.}~\bibnamefont {Liu}}, \bibinfo
  {author} {\bibfnamefont {H.}~\bibnamefont {Lee}}, \bibinfo {author}
  {\bibfnamefont {J.}~\bibnamefont {Dunkel}},  \emph {et~al.},\ }\href@noop {}
  {\bibfield  {journal} {\bibinfo  {journal} {arXiv preprint arXiv:2410.18017}\
  } (\bibinfo {year} {2024})}\BibitemShut {NoStop}%
\bibitem [{\citenamefont {Brandenbourger}\ \emph {et~al.}(2019)\citenamefont
  {Brandenbourger}, \citenamefont {Locsin}, \citenamefont {Lerner},\ and\
  \citenamefont {Coulais}}]{brandenbourger2019non}%
  \BibitemOpen
  \bibfield  {author} {\bibinfo {author} {\bibfnamefont {M.}~\bibnamefont
  {Brandenbourger}}, \bibinfo {author} {\bibfnamefont {X.}~\bibnamefont
  {Locsin}}, \bibinfo {author} {\bibfnamefont {E.}~\bibnamefont {Lerner}}, \
  and\ \bibinfo {author} {\bibfnamefont {C.}~\bibnamefont {Coulais}},\
  }\href@noop {} {\bibfield  {journal} {\bibinfo  {journal} {Nature
  communications}\ }\textbf {\bibinfo {volume} {10}},\ \bibinfo {pages} {4608}
  (\bibinfo {year} {2019})}\BibitemShut {NoStop}%
\bibitem [{\citenamefont {Brandenbourger}\ \emph {et~al.}(2021)\citenamefont
  {Brandenbourger}, \citenamefont {Scheibner}, \citenamefont {Veenstra},
  \citenamefont {Vitelli},\ and\ \citenamefont
  {Coulais}}]{brandenbourger2021limit}%
  \BibitemOpen
  \bibfield  {author} {\bibinfo {author} {\bibfnamefont {M.}~\bibnamefont
  {Brandenbourger}}, \bibinfo {author} {\bibfnamefont {C.}~\bibnamefont
  {Scheibner}}, \bibinfo {author} {\bibfnamefont {J.}~\bibnamefont {Veenstra}},
  \bibinfo {author} {\bibfnamefont {V.}~\bibnamefont {Vitelli}}, \ and\
  \bibinfo {author} {\bibfnamefont {C.}~\bibnamefont {Coulais}},\ }\href@noop
  {} {\bibfield  {journal} {\bibinfo  {journal} {arXiv preprint
  arXiv:2108.08837}\ } (\bibinfo {year} {2021})}\BibitemShut {NoStop}%
\bibitem [{\citenamefont {Veenstra}\ \emph {et~al.}(2024)\citenamefont
  {Veenstra}, \citenamefont {Gamayun}, \citenamefont {Guo}, \citenamefont
  {Sarvi}, \citenamefont {Meinersen},\ and\ \citenamefont
  {Coulais}}]{veenstra2024non}%
  \BibitemOpen
  \bibfield  {author} {\bibinfo {author} {\bibfnamefont {J.}~\bibnamefont
  {Veenstra}}, \bibinfo {author} {\bibfnamefont {O.}~\bibnamefont {Gamayun}},
  \bibinfo {author} {\bibfnamefont {X.}~\bibnamefont {Guo}}, \bibinfo {author}
  {\bibfnamefont {A.}~\bibnamefont {Sarvi}}, \bibinfo {author} {\bibfnamefont
  {C.~V.}\ \bibnamefont {Meinersen}}, \ and\ \bibinfo {author} {\bibfnamefont
  {C.}~\bibnamefont {Coulais}},\ }\href@noop {} {\bibfield  {journal} {\bibinfo
   {journal} {Nature}\ }\textbf {\bibinfo {volume} {627}},\ \bibinfo {pages}
  {528} (\bibinfo {year} {2024})}\BibitemShut {NoStop}%
\bibitem [{\citenamefont {De~Groot}\ and\ \citenamefont
  {Mazur}(2013)}]{de2013non}%
  \BibitemOpen
  \bibfield  {author} {\bibinfo {author} {\bibfnamefont {S.~R.}\ \bibnamefont
  {De~Groot}}\ and\ \bibinfo {author} {\bibfnamefont {P.}~\bibnamefont
  {Mazur}},\ }\href@noop {} {\emph {\bibinfo {title} {Non-equilibrium
  thermodynamics}}}\ (\bibinfo  {publisher} {Courier Corporation},\ \bibinfo
  {year} {2013})\BibitemShut {NoStop}%
\bibitem [{\citenamefont {Chaikin}\ \emph {et~al.}(1995)\citenamefont
  {Chaikin}, \citenamefont {Lubensky},\ and\ \citenamefont
  {Witten}}]{chaikin1995principles}%
  \BibitemOpen
  \bibfield  {author} {\bibinfo {author} {\bibfnamefont {P.~M.}\ \bibnamefont
  {Chaikin}}, \bibinfo {author} {\bibfnamefont {T.~C.}\ \bibnamefont
  {Lubensky}}, \ and\ \bibinfo {author} {\bibfnamefont {T.~A.}\ \bibnamefont
  {Witten}},\ }\href@noop {} {\emph {\bibinfo {title} {Principles of condensed
  matter physics}}},\ Vol.~\bibinfo {volume} {10}\ (\bibinfo  {publisher}
  {Cambridge university press Cambridge},\ \bibinfo {year} {1995})\BibitemShut
  {NoStop}%
\bibitem [{\citenamefont {Prost}\ \emph {et~al.}(2015)\citenamefont {Prost},
  \citenamefont {J{\"u}licher},\ and\ \citenamefont
  {Joanny}}]{prost2015active}%
  \BibitemOpen
  \bibfield  {author} {\bibinfo {author} {\bibfnamefont {J.}~\bibnamefont
  {Prost}}, \bibinfo {author} {\bibfnamefont {F.}~\bibnamefont {J{\"u}licher}},
  \ and\ \bibinfo {author} {\bibfnamefont {J.-F.}\ \bibnamefont {Joanny}},\
  }\href@noop {} {\bibfield  {journal} {\bibinfo  {journal} {Nature physics}\
  }\textbf {\bibinfo {volume} {11}},\ \bibinfo {pages} {111} (\bibinfo {year}
  {2015})}\BibitemShut {NoStop}%
\bibitem [{\citenamefont {Marchetti}\ \emph {et~al.}(2013)\citenamefont
  {Marchetti}, \citenamefont {Joanny}, \citenamefont {Ramaswamy}, \citenamefont
  {Liverpool}, \citenamefont {Prost}, \citenamefont {Rao},\ and\ \citenamefont
  {Simha}}]{marchetti2013hydrodynamics}%
  \BibitemOpen
  \bibfield  {author} {\bibinfo {author} {\bibfnamefont {M.~C.}\ \bibnamefont
  {Marchetti}}, \bibinfo {author} {\bibfnamefont {J.-F.}\ \bibnamefont
  {Joanny}}, \bibinfo {author} {\bibfnamefont {S.}~\bibnamefont {Ramaswamy}},
  \bibinfo {author} {\bibfnamefont {T.~B.}\ \bibnamefont {Liverpool}}, \bibinfo
  {author} {\bibfnamefont {J.}~\bibnamefont {Prost}}, \bibinfo {author}
  {\bibfnamefont {M.}~\bibnamefont {Rao}}, \ and\ \bibinfo {author}
  {\bibfnamefont {R.~A.}\ \bibnamefont {Simha}},\ }\href@noop {} {\bibfield
  {journal} {\bibinfo  {journal} {Reviews of modern physics}\ }\textbf
  {\bibinfo {volume} {85}},\ \bibinfo {pages} {1143} (\bibinfo {year}
  {2013})}\BibitemShut {NoStop}%
\bibitem [{\citenamefont {Cooper}(1996)}]{cooper1996explosives}%
  \BibitemOpen
  \bibfield  {author} {\bibinfo {author} {\bibfnamefont {P.~W.}\ \bibnamefont
  {Cooper}},\ }\href@noop {} {\emph {\bibinfo {title} {Explosives
  engineering}}}\ (\bibinfo  {publisher} {John Wiley \& Sons},\ \bibinfo {year}
  {1996})\BibitemShut {NoStop}%
\bibitem [{\citenamefont {Prigogine}\ and\ \citenamefont
  {Defay}(1958)}]{prigogine1958chemical}%
  \BibitemOpen
  \bibfield  {author} {\bibinfo {author} {\bibfnamefont {I.}~\bibnamefont
  {Prigogine}}\ and\ \bibinfo {author} {\bibfnamefont {R.}~\bibnamefont
  {Defay}},\ }\href@noop {} {\emph {\bibinfo {title} {Chemical
  thermodynamics}}}\ (\bibinfo  {publisher} {Jarrold \& Sons},\ \bibinfo {year}
  {1958})\BibitemShut {NoStop}%
\bibitem [{\citenamefont {Hill}(1986)}]{hill1986introduction}%
  \BibitemOpen
  \bibfield  {author} {\bibinfo {author} {\bibfnamefont {T.~L.}\ \bibnamefont
  {Hill}},\ }\href@noop {} {\emph {\bibinfo {title} {An introduction to
  statistical thermodynamics}}}\ (\bibinfo  {publisher} {Courier Corporation},\
  \bibinfo {year} {1986})\BibitemShut {NoStop}%
\bibitem [{\citenamefont {Hickenboth}\ \emph {et~al.}(2007)\citenamefont
  {Hickenboth}, \citenamefont {Moore}, \citenamefont {White}, \citenamefont
  {Sottos}, \citenamefont {Baudry},\ and\ \citenamefont
  {Wilson}}]{hickenboth2007biasing}%
  \BibitemOpen
  \bibfield  {author} {\bibinfo {author} {\bibfnamefont {C.~R.}\ \bibnamefont
  {Hickenboth}}, \bibinfo {author} {\bibfnamefont {J.~S.}\ \bibnamefont
  {Moore}}, \bibinfo {author} {\bibfnamefont {S.~R.}\ \bibnamefont {White}},
  \bibinfo {author} {\bibfnamefont {N.~R.}\ \bibnamefont {Sottos}}, \bibinfo
  {author} {\bibfnamefont {J.}~\bibnamefont {Baudry}}, \ and\ \bibinfo {author}
  {\bibfnamefont {S.~R.}\ \bibnamefont {Wilson}},\ }\href@noop {} {\bibfield
  {journal} {\bibinfo  {journal} {Nature}\ }\textbf {\bibinfo {volume} {446}},\
  \bibinfo {pages} {423} (\bibinfo {year} {2007})}\BibitemShut {NoStop}%
\bibitem [{\citenamefont {Garcia-Manyes}\ and\ \citenamefont
  {Beedle}(2017)}]{garcia2017steering}%
  \BibitemOpen
  \bibfield  {author} {\bibinfo {author} {\bibfnamefont {S.}~\bibnamefont
  {Garcia-Manyes}}\ and\ \bibinfo {author} {\bibfnamefont {A.~E.}\ \bibnamefont
  {Beedle}},\ }\href@noop {} {\bibfield  {journal} {\bibinfo  {journal} {Nature
  Reviews Chemistry}\ }\textbf {\bibinfo {volume} {1}},\ \bibinfo {pages}
  {0083} (\bibinfo {year} {2017})}\BibitemShut {NoStop}%
\bibitem [{\citenamefont {Ghanem}\ \emph {et~al.}(2021)\citenamefont {Ghanem},
  \citenamefont {Basu}, \citenamefont {Behrou}, \citenamefont {Boechler},
  \citenamefont {Boydston}, \citenamefont {Craig}, \citenamefont {Lin},
  \citenamefont {Lynde}, \citenamefont {Nelson}, \citenamefont {Shen} \emph
  {et~al.}}]{ghanem2021role}%
  \BibitemOpen
  \bibfield  {author} {\bibinfo {author} {\bibfnamefont {M.~A.}\ \bibnamefont
  {Ghanem}}, \bibinfo {author} {\bibfnamefont {A.}~\bibnamefont {Basu}},
  \bibinfo {author} {\bibfnamefont {R.}~\bibnamefont {Behrou}}, \bibinfo
  {author} {\bibfnamefont {N.}~\bibnamefont {Boechler}}, \bibinfo {author}
  {\bibfnamefont {A.~J.}\ \bibnamefont {Boydston}}, \bibinfo {author}
  {\bibfnamefont {S.~L.}\ \bibnamefont {Craig}}, \bibinfo {author}
  {\bibfnamefont {Y.}~\bibnamefont {Lin}}, \bibinfo {author} {\bibfnamefont
  {B.~E.}\ \bibnamefont {Lynde}}, \bibinfo {author} {\bibfnamefont
  {A.}~\bibnamefont {Nelson}}, \bibinfo {author} {\bibfnamefont
  {H.}~\bibnamefont {Shen}},  \emph {et~al.},\ }\href@noop {} {\bibfield
  {journal} {\bibinfo  {journal} {Nature Reviews Materials}\ }\textbf {\bibinfo
  {volume} {6}},\ \bibinfo {pages} {84} (\bibinfo {year} {2021})}\BibitemShut
  {NoStop}%
\bibitem [{\citenamefont {Gilman}(1996)}]{gilman1996mechanochemistry}%
  \BibitemOpen
  \bibfield  {author} {\bibinfo {author} {\bibfnamefont {J.~J.}\ \bibnamefont
  {Gilman}},\ }\href@noop {} {\bibfield  {journal} {\bibinfo  {journal}
  {science}\ }\textbf {\bibinfo {volume} {274}},\ \bibinfo {pages} {65}
  (\bibinfo {year} {1996})}\BibitemShut {NoStop}%
\bibitem [{\citenamefont {O’Neill}\ and\ \citenamefont
  {Boulatov}(2021)}]{o2021many}%
  \BibitemOpen
  \bibfield  {author} {\bibinfo {author} {\bibfnamefont {R.~T.}\ \bibnamefont
  {O’Neill}}\ and\ \bibinfo {author} {\bibfnamefont {R.}~\bibnamefont
  {Boulatov}},\ }\href@noop {} {\bibfield  {journal} {\bibinfo  {journal}
  {Nature Reviews Chemistry}\ }\textbf {\bibinfo {volume} {5}},\ \bibinfo
  {pages} {148} (\bibinfo {year} {2021})}\BibitemShut {NoStop}%
\bibitem [{\citenamefont {Bell}(1978)}]{bell1978models}%
  \BibitemOpen
  \bibfield  {author} {\bibinfo {author} {\bibfnamefont {G.~I.}\ \bibnamefont
  {Bell}},\ }\href@noop {} {\bibfield  {journal} {\bibinfo  {journal}
  {Science}\ }\textbf {\bibinfo {volume} {200}},\ \bibinfo {pages} {618}
  (\bibinfo {year} {1978})}\BibitemShut {NoStop}%
\bibitem [{\citenamefont {Keller}\ and\ \citenamefont
  {Bustamante}(2000)}]{keller2000mechanochemistry}%
  \BibitemOpen
  \bibfield  {author} {\bibinfo {author} {\bibfnamefont {D.}~\bibnamefont
  {Keller}}\ and\ \bibinfo {author} {\bibfnamefont {C.}~\bibnamefont
  {Bustamante}},\ }\href@noop {} {\bibfield  {journal} {\bibinfo  {journal}
  {Biophysical journal}\ }\textbf {\bibinfo {volume} {78}},\ \bibinfo {pages}
  {541} (\bibinfo {year} {2000})}\BibitemShut {NoStop}%
\bibitem [{\citenamefont {Bustamante}\ \emph {et~al.}(2004)\citenamefont
  {Bustamante}, \citenamefont {Chemla}, \citenamefont {Forde},\ and\
  \citenamefont {Izhaky}}]{bustamante2004mechanical}%
  \BibitemOpen
  \bibfield  {author} {\bibinfo {author} {\bibfnamefont {C.}~\bibnamefont
  {Bustamante}}, \bibinfo {author} {\bibfnamefont {Y.~R.}\ \bibnamefont
  {Chemla}}, \bibinfo {author} {\bibfnamefont {N.~R.}\ \bibnamefont {Forde}}, \
  and\ \bibinfo {author} {\bibfnamefont {D.}~\bibnamefont {Izhaky}},\
  }\href@noop {} {\bibfield  {journal} {\bibinfo  {journal} {Annual review of
  biochemistry}\ }\textbf {\bibinfo {volume} {73}},\ \bibinfo {pages} {705}
  (\bibinfo {year} {2004})}\BibitemShut {NoStop}%
\bibitem [{\citenamefont {Vogel}(2018)}]{vogel2018unraveling}%
  \BibitemOpen
  \bibfield  {author} {\bibinfo {author} {\bibfnamefont {V.}~\bibnamefont
  {Vogel}},\ }\href@noop {} {\bibfield  {journal} {\bibinfo  {journal} {Annual
  review of physiology}\ }\textbf {\bibinfo {volume} {80}},\ \bibinfo {pages}
  {353} (\bibinfo {year} {2018})}\BibitemShut {NoStop}%
\bibitem [{\citenamefont {Bailles}\ \emph {et~al.}(2022)\citenamefont
  {Bailles}, \citenamefont {Gehrels},\ and\ \citenamefont
  {Lecuit}}]{bailles2022mechanochemical}%
  \BibitemOpen
  \bibfield  {author} {\bibinfo {author} {\bibfnamefont {A.}~\bibnamefont
  {Bailles}}, \bibinfo {author} {\bibfnamefont {E.~W.}\ \bibnamefont
  {Gehrels}}, \ and\ \bibinfo {author} {\bibfnamefont {T.}~\bibnamefont
  {Lecuit}},\ }\href@noop {} {\bibfield  {journal} {\bibinfo  {journal} {Annual
  review of cell and developmental biology}\ }\textbf {\bibinfo {volume}
  {38}},\ \bibinfo {pages} {321} (\bibinfo {year} {2022})}\BibitemShut
  {NoStop}%
\bibitem [{\citenamefont {H{\"a}nggi}\ \emph {et~al.}(1990)\citenamefont
  {H{\"a}nggi}, \citenamefont {Talkner},\ and\ \citenamefont
  {Borkovec}}]{hanggi1990reaction}%
  \BibitemOpen
  \bibfield  {author} {\bibinfo {author} {\bibfnamefont {P.}~\bibnamefont
  {H{\"a}nggi}}, \bibinfo {author} {\bibfnamefont {P.}~\bibnamefont {Talkner}},
  \ and\ \bibinfo {author} {\bibfnamefont {M.}~\bibnamefont {Borkovec}},\
  }\href@noop {} {\bibfield  {journal} {\bibinfo  {journal} {Reviews of modern
  physics}\ }\textbf {\bibinfo {volume} {62}},\ \bibinfo {pages} {251}
  (\bibinfo {year} {1990})}\BibitemShut {NoStop}%
\bibitem [{\citenamefont {Zwicker}(2022)}]{zwicker2022intertwined}%
  \BibitemOpen
  \bibfield  {author} {\bibinfo {author} {\bibfnamefont {D.}~\bibnamefont
  {Zwicker}},\ }\href@noop {} {\bibfield  {journal} {\bibinfo  {journal}
  {Current Opinion in Colloid \& Interface Science}\ }\textbf {\bibinfo
  {volume} {61}},\ \bibinfo {pages} {101606} (\bibinfo {year}
  {2022})}\BibitemShut {NoStop}%
\bibitem [{\citenamefont {Aslyamov}\ \emph {et~al.}(2023)\citenamefont
  {Aslyamov}, \citenamefont {Avanzini}, \citenamefont {Fodor},\ and\
  \citenamefont {Esposito}}]{aslyamov2023nonideal}%
  \BibitemOpen
  \bibfield  {author} {\bibinfo {author} {\bibfnamefont {T.}~\bibnamefont
  {Aslyamov}}, \bibinfo {author} {\bibfnamefont {F.}~\bibnamefont {Avanzini}},
  \bibinfo {author} {\bibfnamefont {{\'E}.}~\bibnamefont {Fodor}}, \ and\
  \bibinfo {author} {\bibfnamefont {M.}~\bibnamefont {Esposito}},\ }\href@noop
  {} {\bibfield  {journal} {\bibinfo  {journal} {Physical Review Letters}\
  }\textbf {\bibinfo {volume} {131}},\ \bibinfo {pages} {138301} (\bibinfo
  {year} {2023})}\BibitemShut {NoStop}%
\bibitem [{SM2()}]{SM2024}%
  \BibitemOpen
  \href@noop {} {\enquote {\bibinfo {title} {See \uppercase{S}upplemental
  \uppercase{M}aterial ... for details},}\ }\BibitemShut {NoStop}%
\bibitem [{\citenamefont {Orszag}(1972)}]{orszag1972comparison}%
  \BibitemOpen
  \bibfield  {author} {\bibinfo {author} {\bibfnamefont {S.~A.}\ \bibnamefont
  {Orszag}},\ }\href@noop {} {\bibfield  {journal} {\bibinfo  {journal}
  {Studies in Applied Mathematics}\ }\textbf {\bibinfo {volume} {51}},\
  \bibinfo {pages} {253} (\bibinfo {year} {1972})}\BibitemShut {NoStop}%
\bibitem [{\citenamefont {You}\ \emph {et~al.}(2020)\citenamefont {You},
  \citenamefont {Baskaran},\ and\ \citenamefont
  {Marchetti}}]{you2020nonreciprocity}%
  \BibitemOpen
  \bibfield  {author} {\bibinfo {author} {\bibfnamefont {Z.}~\bibnamefont
  {You}}, \bibinfo {author} {\bibfnamefont {A.}~\bibnamefont {Baskaran}}, \
  and\ \bibinfo {author} {\bibfnamefont {M.~C.}\ \bibnamefont {Marchetti}},\
  }\href@noop {} {\bibfield  {journal} {\bibinfo  {journal} {Proceedings of the
  National Academy of Sciences}\ }\textbf {\bibinfo {volume} {117}},\ \bibinfo
  {pages} {19767} (\bibinfo {year} {2020})}\BibitemShut {NoStop}%
\bibitem [{\citenamefont {Saha}\ \emph {et~al.}(2020)\citenamefont {Saha},
  \citenamefont {Agudo-Canalejo},\ and\ \citenamefont
  {Golestanian}}]{saha2020scalar}%
  \BibitemOpen
  \bibfield  {author} {\bibinfo {author} {\bibfnamefont {S.}~\bibnamefont
  {Saha}}, \bibinfo {author} {\bibfnamefont {J.}~\bibnamefont
  {Agudo-Canalejo}}, \ and\ \bibinfo {author} {\bibfnamefont {R.}~\bibnamefont
  {Golestanian}},\ }\href@noop {} {\bibfield  {journal} {\bibinfo  {journal}
  {Physical Review X}\ }\textbf {\bibinfo {volume} {10}},\ \bibinfo {pages}
  {041009} (\bibinfo {year} {2020})}\BibitemShut {NoStop}%
\bibitem [{\citenamefont {Brauns}\ and\ \citenamefont
  {Marchetti}(2024)}]{brauns2024nonreciprocal}%
  \BibitemOpen
  \bibfield  {author} {\bibinfo {author} {\bibfnamefont {F.}~\bibnamefont
  {Brauns}}\ and\ \bibinfo {author} {\bibfnamefont {M.~C.}\ \bibnamefont
  {Marchetti}},\ }\href@noop {} {\bibfield  {journal} {\bibinfo  {journal}
  {Physical Review X}\ }\textbf {\bibinfo {volume} {14}},\ \bibinfo {pages}
  {021014} (\bibinfo {year} {2024})}\BibitemShut {NoStop}%
\bibitem [{\citenamefont {Stanton}\ \emph {et~al.}(2012)\citenamefont
  {Stanton}, \citenamefont {Owens},\ and\ \citenamefont
  {Mann}}]{STANTON20123617}%
  \BibitemOpen
  \bibfield  {author} {\bibinfo {author} {\bibfnamefont {S.~C.}\ \bibnamefont
  {Stanton}}, \bibinfo {author} {\bibfnamefont {B.~A.}\ \bibnamefont {Owens}},
  \ and\ \bibinfo {author} {\bibfnamefont {B.~P.}\ \bibnamefont {Mann}},\
  }\href {\doibase https://doi.org/10.1016/j.jsv.2012.03.012} {\bibfield
  {journal} {\bibinfo  {journal} {Journal of Sound and Vibration}\ }\textbf
  {\bibinfo {volume} {331}},\ \bibinfo {pages} {3617} (\bibinfo {year}
  {2012})}\BibitemShut {NoStop}%
\bibitem [{\citenamefont {Johansson}\ \emph {et~al.}(2014)\citenamefont
  {Johansson}, \citenamefont {Prilepsky},\ and\ \citenamefont
  {Derevyanko}}]{johansson2014strongly}%
  \BibitemOpen
  \bibfield  {author} {\bibinfo {author} {\bibfnamefont {M.}~\bibnamefont
  {Johansson}}, \bibinfo {author} {\bibfnamefont {J.~E.}\ \bibnamefont
  {Prilepsky}}, \ and\ \bibinfo {author} {\bibfnamefont {S.~A.}\ \bibnamefont
  {Derevyanko}},\ }\href@noop {} {\bibfield  {journal} {\bibinfo  {journal}
  {Physical Review E}\ }\textbf {\bibinfo {volume} {89}},\ \bibinfo {pages}
  {042912} (\bibinfo {year} {2014})}\BibitemShut {NoStop}%
\bibitem [{\citenamefont {Mohammadpour}\ \emph {et~al.}(2023)\citenamefont
  {Mohammadpour}, \citenamefont {Abdelkefi}, \citenamefont {Safarpour},
  \citenamefont {Gavagsaz-Ghoachani},\ and\ \citenamefont
  {Zandi}}]{mohammadpour2023controlling}%
  \BibitemOpen
  \bibfield  {author} {\bibinfo {author} {\bibfnamefont {M.}~\bibnamefont
  {Mohammadpour}}, \bibinfo {author} {\bibfnamefont {A.}~\bibnamefont
  {Abdelkefi}}, \bibinfo {author} {\bibfnamefont {P.}~\bibnamefont
  {Safarpour}}, \bibinfo {author} {\bibfnamefont {R.}~\bibnamefont
  {Gavagsaz-Ghoachani}}, \ and\ \bibinfo {author} {\bibfnamefont
  {M.}~\bibnamefont {Zandi}},\ }\href@noop {} {\bibfield  {journal} {\bibinfo
  {journal} {Meccanica}\ }\textbf {\bibinfo {volume} {58}},\ \bibinfo {pages}
  {587} (\bibinfo {year} {2023})}\BibitemShut {NoStop}%
\bibitem [{\citenamefont {Erturk}\ and\ \citenamefont
  {Inman}(2011)}]{erturk2011broadband}%
  \BibitemOpen
  \bibfield  {author} {\bibinfo {author} {\bibfnamefont {A.}~\bibnamefont
  {Erturk}}\ and\ \bibinfo {author} {\bibfnamefont {D.~J.}\ \bibnamefont
  {Inman}},\ }\href@noop {} {\bibfield  {journal} {\bibinfo  {journal} {Journal
  of Sound and Vibration}\ }\textbf {\bibinfo {volume} {330}},\ \bibinfo
  {pages} {2339} (\bibinfo {year} {2011})}\BibitemShut {NoStop}%
\bibitem [{\citenamefont {Konotop}\ \emph {et~al.}(2016)\citenamefont
  {Konotop}, \citenamefont {Yang},\ and\ \citenamefont
  {Zezyulin}}]{RevModPhys.88.035002}%
  \BibitemOpen
  \bibfield  {author} {\bibinfo {author} {\bibfnamefont {V.~V.}\ \bibnamefont
  {Konotop}}, \bibinfo {author} {\bibfnamefont {J.}~\bibnamefont {Yang}}, \
  and\ \bibinfo {author} {\bibfnamefont {D.~A.}\ \bibnamefont {Zezyulin}},\
  }\href {\doibase 10.1103/RevModPhys.88.035002} {\bibfield  {journal}
  {\bibinfo  {journal} {Rev. Mod. Phys.}\ }\textbf {\bibinfo {volume} {88}},\
  \bibinfo {pages} {035002} (\bibinfo {year} {2016})}\BibitemShut {NoStop}%
\bibitem [{\citenamefont {Zhang}\ \emph {et~al.}(2015)\citenamefont {Zhang},
  \citenamefont {Peng}, \citenamefont {\"Ozdemir}, \citenamefont {Liu},
  \citenamefont {Jing}, \citenamefont {L\"u}, \citenamefont {Liu},
  \citenamefont {Yang},\ and\ \citenamefont {Nori}}]{PhysRevB.92.115407}%
  \BibitemOpen
  \bibfield  {author} {\bibinfo {author} {\bibfnamefont {J.}~\bibnamefont
  {Zhang}}, \bibinfo {author} {\bibfnamefont {B.}~\bibnamefont {Peng}},
  \bibinfo {author} {\bibfnamefont {i.~m. c.~K.}\ \bibnamefont {\"Ozdemir}},
  \bibinfo {author} {\bibfnamefont {Y.-x.}\ \bibnamefont {Liu}}, \bibinfo
  {author} {\bibfnamefont {H.}~\bibnamefont {Jing}}, \bibinfo {author}
  {\bibfnamefont {X.-y.}\ \bibnamefont {L\"u}}, \bibinfo {author}
  {\bibfnamefont {Y.-l.}\ \bibnamefont {Liu}}, \bibinfo {author} {\bibfnamefont
  {L.}~\bibnamefont {Yang}}, \ and\ \bibinfo {author} {\bibfnamefont
  {F.}~\bibnamefont {Nori}},\ }\href {\doibase 10.1103/PhysRevB.92.115407}
  {\bibfield  {journal} {\bibinfo  {journal} {Phys. Rev. B}\ }\textbf {\bibinfo
  {volume} {92}},\ \bibinfo {pages} {115407} (\bibinfo {year}
  {2015})}\BibitemShut {NoStop}%
\bibitem [{\citenamefont {Browning}\ \emph {et~al.}(2019)\citenamefont
  {Browning}, \citenamefont {Woodhouse},\ and\ \citenamefont
  {Simpson}}]{browning2019reversible}%
  \BibitemOpen
  \bibfield  {author} {\bibinfo {author} {\bibfnamefont {A.~P.}\ \bibnamefont
  {Browning}}, \bibinfo {author} {\bibfnamefont {F.~G.}\ \bibnamefont
  {Woodhouse}}, \ and\ \bibinfo {author} {\bibfnamefont {M.~J.}\ \bibnamefont
  {Simpson}},\ }\href@noop {} {\bibfield  {journal} {\bibinfo  {journal}
  {Proceedings of the Royal Society A}\ }\textbf {\bibinfo {volume} {475}},\
  \bibinfo {pages} {20190146} (\bibinfo {year} {2019})}\BibitemShut {NoStop}%
\bibitem [{\citenamefont {Reichenbach}\ and\ \citenamefont
  {Hudspeth}(2014)}]{reichenbach2014physics}%
  \BibitemOpen
  \bibfield  {author} {\bibinfo {author} {\bibfnamefont {T.}~\bibnamefont
  {Reichenbach}}\ and\ \bibinfo {author} {\bibfnamefont {A.}~\bibnamefont
  {Hudspeth}},\ }\href@noop {} {\bibfield  {journal} {\bibinfo  {journal}
  {Reports on Progress in Physics}\ }\textbf {\bibinfo {volume} {77}},\
  \bibinfo {pages} {076601} (\bibinfo {year} {2014})}\BibitemShut {NoStop}%
\bibitem [{\citenamefont {Martin}\ and\ \citenamefont
  {Hudspeth}(2021)}]{martin2021mechanical}%
  \BibitemOpen
  \bibfield  {author} {\bibinfo {author} {\bibfnamefont {P.}~\bibnamefont
  {Martin}}\ and\ \bibinfo {author} {\bibfnamefont {A.}~\bibnamefont
  {Hudspeth}},\ }\href@noop {} {\bibfield  {journal} {\bibinfo  {journal}
  {Annual Review of Condensed Matter Physics}\ }\textbf {\bibinfo {volume}
  {12}},\ \bibinfo {pages} {29} (\bibinfo {year} {2021})}\BibitemShut {NoStop}%
\bibitem [{\citenamefont {Dhooge}\ \emph {et~al.}(2003)\citenamefont {Dhooge},
  \citenamefont {Govaerts},\ and\ \citenamefont {Kuznetsov}}]{matcont}%
  \BibitemOpen
  \bibfield  {author} {\bibinfo {author} {\bibfnamefont {A.}~\bibnamefont
  {Dhooge}}, \bibinfo {author} {\bibfnamefont {W.}~\bibnamefont {Govaerts}}, \
  and\ \bibinfo {author} {\bibfnamefont {Y.~A.}\ \bibnamefont {Kuznetsov}},\
  }\href {\doibase 10.1145/779359.779362} {\bibfield  {journal} {\bibinfo
  {journal} {ACM Trans. Math. Softw.}\ }\textbf {\bibinfo {volume} {29}},\
  \bibinfo {pages} {141–164} (\bibinfo {year} {2003})}\BibitemShut {NoStop}%
\bibitem [{\citenamefont {Scheibner}\ \emph
  {et~al.}(2020{\natexlab{b}})\citenamefont {Scheibner}, \citenamefont
  {Irvine},\ and\ \citenamefont {Vitelli}}]{scheibner2020non}%
  \BibitemOpen
  \bibfield  {author} {\bibinfo {author} {\bibfnamefont {C.}~\bibnamefont
  {Scheibner}}, \bibinfo {author} {\bibfnamefont {W.~T.}\ \bibnamefont
  {Irvine}}, \ and\ \bibinfo {author} {\bibfnamefont {V.}~\bibnamefont
  {Vitelli}},\ }\href@noop {} {\bibfield  {journal} {\bibinfo  {journal}
  {Physical Review Letters}\ }\textbf {\bibinfo {volume} {125}},\ \bibinfo
  {pages} {118001} (\bibinfo {year} {2020}{\natexlab{b}})}\BibitemShut
  {NoStop}%
\bibitem [{\citenamefont {Strogatz}(2024)}]{strogatz2024nonlinear}%
  \BibitemOpen
  \bibfield  {author} {\bibinfo {author} {\bibfnamefont {S.~H.}\ \bibnamefont
  {Strogatz}},\ }\href@noop {} {\emph {\bibinfo {title} {Nonlinear dynamics and
  chaos: with applications to physics, biology, chemistry, and engineering}}}\
  (\bibinfo  {publisher} {Chapman and Hall/CRC},\ \bibinfo {year}
  {2024})\BibitemShut {NoStop}%
\bibitem [{\citenamefont {Elliott}\ and\ \citenamefont
  {Leys}(2007)}]{elliott2007coordinated}%
  \BibitemOpen
  \bibfield  {author} {\bibinfo {author} {\bibfnamefont {G.~R.}\ \bibnamefont
  {Elliott}}\ and\ \bibinfo {author} {\bibfnamefont {S.~P.}\ \bibnamefont
  {Leys}},\ }\href@noop {} {\bibfield  {journal} {\bibinfo  {journal} {Journal
  of Experimental Biology}\ }\textbf {\bibinfo {volume} {210}},\ \bibinfo
  {pages} {3736} (\bibinfo {year} {2007})}\BibitemShut {NoStop}%
\bibitem [{\citenamefont {Yuan}\ \emph {et~al.}(2017)\citenamefont {Yuan},
  \citenamefont {McCracken}, \citenamefont {Gross}, \citenamefont {Braun},
  \citenamefont {Moore},\ and\ \citenamefont {Nuzzo}}]{yuan_programmable}%
  \BibitemOpen
  \bibfield  {author} {\bibinfo {author} {\bibfnamefont {P.}~\bibnamefont
  {Yuan}}, \bibinfo {author} {\bibfnamefont {J.~M.}\ \bibnamefont {McCracken}},
  \bibinfo {author} {\bibfnamefont {D.~E.}\ \bibnamefont {Gross}}, \bibinfo
  {author} {\bibfnamefont {P.~V.}\ \bibnamefont {Braun}}, \bibinfo {author}
  {\bibfnamefont {J.~S.}\ \bibnamefont {Moore}}, \ and\ \bibinfo {author}
  {\bibfnamefont {R.~G.}\ \bibnamefont {Nuzzo}},\ }\href@noop {} {\bibfield
  {journal} {\bibinfo  {journal} {Soft Matter}\ }\textbf {\bibinfo {volume}
  {13}},\ \bibinfo {pages} {7312} (\bibinfo {year} {2017})}\BibitemShut
  {NoStop}%
\bibitem [{\citenamefont {Metze}\ \emph {et~al.}(2023)\citenamefont {Metze},
  \citenamefont {Sant}, \citenamefont {Meng}, \citenamefont {Klok},\ and\
  \citenamefont {Kaur}}]{metze2023swelling}%
  \BibitemOpen
  \bibfield  {author} {\bibinfo {author} {\bibfnamefont {F.~K.}\ \bibnamefont
  {Metze}}, \bibinfo {author} {\bibfnamefont {S.}~\bibnamefont {Sant}},
  \bibinfo {author} {\bibfnamefont {Z.}~\bibnamefont {Meng}}, \bibinfo {author}
  {\bibfnamefont {H.-A.}\ \bibnamefont {Klok}}, \ and\ \bibinfo {author}
  {\bibfnamefont {K.}~\bibnamefont {Kaur}},\ }\href@noop {} {\bibfield
  {journal} {\bibinfo  {journal} {Langmuir}\ }\textbf {\bibinfo {volume}
  {39}},\ \bibinfo {pages} {3546} (\bibinfo {year} {2023})}\BibitemShut
  {NoStop}%
\bibitem [{\citenamefont {Ouchi}\ \emph {et~al.}(2023)\citenamefont {Ouchi},
  \citenamefont {Bowser}, \citenamefont {Kouznetsova}, \citenamefont {Zheng},\
  and\ \citenamefont {Craig}}]{ouchi2023strain}%
  \BibitemOpen
  \bibfield  {author} {\bibinfo {author} {\bibfnamefont {T.}~\bibnamefont
  {Ouchi}}, \bibinfo {author} {\bibfnamefont {B.~H.}\ \bibnamefont {Bowser}},
  \bibinfo {author} {\bibfnamefont {T.~B.}\ \bibnamefont {Kouznetsova}},
  \bibinfo {author} {\bibfnamefont {X.}~\bibnamefont {Zheng}}, \ and\ \bibinfo
  {author} {\bibfnamefont {S.~L.}\ \bibnamefont {Craig}},\ }\href@noop {}
  {\bibfield  {journal} {\bibinfo  {journal} {Materials Horizons}\ }\textbf
  {\bibinfo {volume} {10}},\ \bibinfo {pages} {585} (\bibinfo {year}
  {2023})}\BibitemShut {NoStop}%
\bibitem [{\citenamefont {Kim}\ \emph {et~al.}(2021)\citenamefont {Kim},
  \citenamefont {Zhang}, \citenamefont {Shi},\ and\ \citenamefont
  {Suo}}]{kim2021fracture}%
  \BibitemOpen
  \bibfield  {author} {\bibinfo {author} {\bibfnamefont {J.}~\bibnamefont
  {Kim}}, \bibinfo {author} {\bibfnamefont {G.}~\bibnamefont {Zhang}}, \bibinfo
  {author} {\bibfnamefont {M.}~\bibnamefont {Shi}}, \ and\ \bibinfo {author}
  {\bibfnamefont {Z.}~\bibnamefont {Suo}},\ }\href@noop {} {\bibfield
  {journal} {\bibinfo  {journal} {Science}\ }\textbf {\bibinfo {volume}
  {374}},\ \bibinfo {pages} {212} (\bibinfo {year} {2021})}\BibitemShut
  {NoStop}%
\bibitem [{\citenamefont {Westrin}\ \emph {et~al.}(1994)\citenamefont
  {Westrin}, \citenamefont {Axelsson},\ and\ \citenamefont
  {Zacchi}}]{westrin1994diffusion}%
  \BibitemOpen
  \bibfield  {author} {\bibinfo {author} {\bibfnamefont {B.~A.}\ \bibnamefont
  {Westrin}}, \bibinfo {author} {\bibfnamefont {A.}~\bibnamefont {Axelsson}}, \
  and\ \bibinfo {author} {\bibfnamefont {G.}~\bibnamefont {Zacchi}},\
  }\href@noop {} {\bibfield  {journal} {\bibinfo  {journal} {Journal of
  Controlled Release}\ }\textbf {\bibinfo {volume} {30}},\ \bibinfo {pages}
  {189} (\bibinfo {year} {1994})}\BibitemShut {NoStop}%
\bibitem [{\citenamefont {Joanny}\ and\ \citenamefont
  {Prost}(2009)}]{joanny2009active}%
  \BibitemOpen
  \bibfield  {author} {\bibinfo {author} {\bibfnamefont {J.-F.}\ \bibnamefont
  {Joanny}}\ and\ \bibinfo {author} {\bibfnamefont {J.}~\bibnamefont {Prost}},\
  }\href@noop {} {\bibfield  {journal} {\bibinfo  {journal} {HFSP journal}\
  }\textbf {\bibinfo {volume} {3}},\ \bibinfo {pages} {94} (\bibinfo {year}
  {2009})}\BibitemShut {NoStop}%
\bibitem [{\citenamefont {Yang}\ \emph {et~al.}(2020)\citenamefont {Yang},
  \citenamefont {Chang},\ and\ \citenamefont
  {P{\'e}rez-Arancibia}}]{yang202088}%
  \BibitemOpen
  \bibfield  {author} {\bibinfo {author} {\bibfnamefont {X.}~\bibnamefont
  {Yang}}, \bibinfo {author} {\bibfnamefont {L.}~\bibnamefont {Chang}}, \ and\
  \bibinfo {author} {\bibfnamefont {N.~O.}\ \bibnamefont
  {P{\'e}rez-Arancibia}},\ }\href@noop {} {\bibfield  {journal} {\bibinfo
  {journal} {Science Robotics}\ }\textbf {\bibinfo {volume} {5}},\ \bibinfo
  {pages} {eaba0015} (\bibinfo {year} {2020})}\BibitemShut {NoStop}%
\end{thebibliography}

%

\end{document}